# Wall boundary conditions for Lattice Boltzmann simulations of turbulent flows with wall functions

**Jorge Ponsin*, Carlos Lozano**

*Theoretical and Computational Aerodynamics*

*National Institute of Aerospace Technology (INTA)*

*Torrejón de Ardoz 28850, Madrid, Spain*

**Corresponding author: ponsinj@inta.es*

## Summary

This paper investigates wall boundary condition schemes for the simulation of turbulent flows using the Lattice Boltzmann method (LBM) coupled to turbulence models with wall functions. The analysis focuses on two schemes: a regularized boundary scheme with third-order reconstruction of the velocity gradients using wall function data and a slip-velocity bounce-back scheme. The LBM solver is coupled to the Spalart-Allmaras turbulence model and uses a model consistent wall function. The performance of the wall boundary schemes is assessed in two canonical turbulent flow cases, a fully developed channel flow and a zero-pressure-gradient flat plate boundary layer (BL), selected specifically to isolate and analyze the impact of wall boundary treatments on turbulence modeling. The analysis shows that, for the selected test cases, the slip-velocity bounce-back approach, which has received relatively little attention within the context of LBM coupled to Reynolds-Averaged Navier-Stokes (RANS) turbulence models with wall functions, behaves fairly consistently in terms of both accuracy and mesh convergence. The regularized-based approach, on the other hand, appears to be highly sensitive to the reconstruction of the wall-normal velocity gradient, even in simple geometries such as flat walls where no interpolation is required. This dependency of the regularized boundary schemes on near-wall gradients, which had been noted before in the literature, requires the use of ad-hoc gradient reconstruction techniques, requirements that are not present in the slip-velocity bounce-back method. A hybrid regularized boundary scheme that blends two different gradient reconstruction techniques but requires calibration is introduced as a tool to investigate this effect.

## 1. Introduction

The application of high-fidelity Computational Fluid Dynamics (CFD) methods for predicting complex physical flow scenarios involving unsteady turbulent flows has gained significant attention over the past two decades. In external aerodynamics, notable examples include the characterization of flow separation in high-lift configurations [1], the wing-body juncture flow [2], buffet prediction [3], and acoustic noise prediction [4], among others. Methods based on the RANS equations have shown limited predictive capabilities in these cases. Most scale-resolving CFD approaches for high-Reynolds-number turbulent flows rely on hybrid RANS/LES methods [5]. In these approaches, a Large Eddy Simulation (LES) model is used to resolve the energetic turbulent scales in regions away from the wall, while RANS equations are applied near the wall. The activation of the LES and RANS models can be determined implicitly through turbulence model equations, as in the Detached Eddy Simulation (DES) approach and its variants [6], or by explicitly partitioning the computational domain into LES and RANS zones, which requires special care in handling the interface between both zones. One example of this zonal approach is the *overset-LES* technique, which has been applied in noise prediction studies [7, 8, 9]. Another popular approach is Wall-Modeled LES (WMLES), in which the LES method is applied at the outer part of the boundary layer, while the near wall region is modeled using simplified wall models to reduce the computational cost [10].

Scale-resolving methods typically require specialized numerical techniques with low numerical dissipation, such as high-order Finite Volume, Finite Element, and Discontinuous Garlerkin Methods, etc. On the other hand, the small physical time steps required to resolve turbulent structures, combined with the grid requirements in the LES region, make scale-resolving methods computationally expensive within the standard CFD framework. In this context, the Lattice Boltzmann Method (LBM) has emerged as an alternative approach for unsteady and/or scale-resolving turbulent flow applications [11, 12, 13, 14] due to several advantageous characteristics. First, it is intrinsically formulated for unsteady flows and has very low numerical dissipation/dispersion errors, which makes it suitable for joint use with high-fidelity turbulence simulation approaches such as Large-eddy simulation (LES). Second, its underlying adaptive Cartesian octree mesh approach, together with the way wall boundary conditions are imposed, is particularly well-suited to deal with complex and/or moving geometries. Finally, LBM is easily parallelizable due to the explicit and compact nature of the stream-collide algorithm which can be efficiently adapted to all kinds of parallel architectures (either CPU or GPU-based).

The LBM solves the unsteady Navier-Stokes equations and is, thus, in this regard, equivalent to direct numerical simulation (DNS). In order to simulate high Reynolds number flows without incurring excessive computational cost, LBM must be supplemented with either a RANS turbulence model (such as the Spalart-Allmaras model) or a subgrid scale model which models only the smallest scales than cannot be adequately simulated with the computational mesh. For conciseness, we will refer to the former as LBM-RANS and the latter as LBM-LES, but the reader is warned that this nomenclature does not imply that LBM-RANS models solve the RANS equations, but rather that LBM computations converge to RANS-like solutions statistically.

Since standard LBM relies on isotropic Cartesian grids, resolving turbulent boundary layers at high Reynolds numbers demands extremely fine meshes, making those simulations computationally expensive or impractical. To address this issue, wall functions are widely used in both RANS and WMLES-based LBM simulations. However, implementing accurate wall boundary conditions for high-Reynolds turbulent flows remains an open challenge in LBM. While several alternatives are available in the specialized literature, none has demonstrated a superior performance so as to become a standard. There are at least two reasons for that. On one side, boundary conditions in LBM can only be imposed via suitable boundary population distributions, whose number is usually greater than the number of macroscopic physical conditions to be imposed. Hence, there is considerable freedom in prescribing those populations. On the other side, and connected with the previous statement, it is inherently difficult in LBM to link the definition of the mesoscopic population distributions at boundary nodes with the physical (macroscopic) boundary conditions, especially when wall functions are used.

Wall boundary conditions within the LBM framework fall essentially into two different categories: the bounce-back-based schemes [15] and the regularized boundary condition methods [16]. Originally developed for laminar flow simulations, both techniques have subsequently been extended and adapted to handle turbulent flows, particularly in the context of RANS and WMLES models [17, 18, 19, 20, 21, 22]. The bounce-back method is a link-wise boundary scheme [23] in which post-collision distribution functions that propagate towards a solid boundary are reflected in the opposite direction, effectively reversing their velocity vectors. This reflection enforces the no-slip condition by conserving momentum at the boundary node, where only incoming distribution functions are prescribed. The bounce-back scheme attains second-order accuracy when the boundary node is positioned exactly halfway between fluid and solid nodes [24]. For arbitrarily curved geometries, the placement of boundary nodes at non-ideal locations is unavoidable. To address this problem, link-wise interpolated bounce-back schemes have been developed [25] [26], allowing for improved spatial accuracy. These schemes have been shown to recover second-order accuracy for simple curved boundaries in laminar regimes.

The application of the bounce-back scheme in turbulent flow simulations using wall functions within the LBM framework is significantly less explored compared to its extensive use in laminar flow contexts. This is certainly the case with LBM-RANS simulations with wall functions where, to the best of our knowledge, there is currently no published research that directly investigates the implementation of bounce-back schemes. On the other hand, preliminary efforts have been made within the framework of LBM coupled to a WMLES model using shear-stress-based equilibrium wall models. For instance, Pasquali [27] proposed a bounce-back scheme integrated with an implicit WMLES formulation and a cumulant-based collision operator, validating the methodology on a turbulent channel flow. However, this work focuses primarily on the role of the numerical dissipation introduced by the collision operator rather than on a detailed analysis of the wall treatment itself. Nishimura [28] considered a shear-stress-based implicit WMLES framework with wall functions that incorporates Yu's link-wise interpolated bounce-back scheme [26] to handle complex geometries with curved surfaces. Their focus lies in the simulation of aero-acoustic phenomena, particularly in multi-element configurations, with validation efforts centered on acoustic spectra prediction. In both cases, the central objective is the application of high-fidelity turbulence modeling in conjunction with the LBM framework. As such, the impact of the wall boundary condition cannot be isolated, making it difficult to draw definitive conclusions regarding the bounce-back scheme's standalone performance in turbulent flow regimes.

Unlike the bounce-back method, the regularized wall boundary condition reconstructs the full set of populations at the boundary node, including those streamed from the interior. This method is inspired on the Chapman-Enskog (CE) expansion [16] and treats differently the equilibrium and non-equilibrium parts of the distribution functions. The equilibrium component, that depends on macroscopic variables (density and velocity), is obtained via interpolation/extrapolation from interior nodes [17], while the non-equilibrium part is reconstructed (or *regularized*) using the first-order CE term, which can be computed either mesoscopically from the second-order moments of the populations [16] [18] or from the macroscopic strain rate tensor, which can be used for general geometries but requires reconstruction of the velocity gradients using finite-difference schemes [16] [17] [29].

Regularized wall boundary conditions are more widely used than bounce-back methods for LBM simulations of turbulent flows. Malaspinas et al. [18] applied a mesoscopic-based regularized approach within WMLES and equilibrium wall functions to simulate turbulent channel flows across various friction Reynolds numbers, achieving strong agreement with reference data. Willhem et al. employed a hybrid regularized scheme using finite-differences reconstruction of non-equilibrium populations in LBM-RANS simulations of airfoils [19] and WMLES of wings [30]. Cai and Degrigny [20] [21] enhanced the application of the regularized approach to LBM-RANS by introducing techniques for handling immersed boundaries, including a novel method to compute normal velocity gradients using finite differences applied to the analytical wall function. They validated their approach on flat plate and NACA 0012 airfoil test cases. Recently, Lyu et al. [31] employed the Willhelm approach within a hybrid method for local wall refinement and applied it to the prediction of a pair of counter-rotating propellers using an LBM-RANS-based approach. Haussman et al. [22] [32] proposed another variant, where non-equilibrium populations were reconstructed by extrapolation from interior nodes [33], and validated it for turbulent channel flows using WMLES with wall functions.

Finally, it is worth noting that alternative LBM wall boundary methods, such as the ghost-cell and volumetric approaches, have also been applied to turbulent flows. The ghost-cell method [34] reconstructs equilibrium and non-equilibrium populations using extrapolation/interpolation techniques based on ghost points. Pellerin [34] applied this approach to LBM-RANS simulations without wall modeling, resolving turbulence up to the wall and enforcing no-slip conditions on curved surfaces. The volumetric formulation [35], on the other hand, enforces wall boundary conditions conservatively by using auxiliary geometric elements (facets and volumetric regions) and a precise algorithm that blends free-slip and non-slip treatments to prescribe the turbulent wall flux obtained from a wall function. This approach, implemented in the commercial LBM solver *PowerFLOW©*, has been successfully used for VLES simulations of complex geometries [14].

The above literature review unveils a significant gap in the evaluation of LBM boundary conditions for turbulent flows with wall functions. Existing studies often involve a wide range of modeling approaches, turbulence models and geometric complexities, making it challenging to isolate and quantify the influence of the boundary condition scheme on the overall accuracy of the solution. Consequently, the effect of wall boundary treatments remains underexplored and poorly understood in the context of turbulent LBM simulations. The primary objective of this work is to address this gap by systematically investigating and comparing different LBM wall boundary condition schemes for RANS-based simulations. To the best of the authors' knowledge, this is the first study to directly compare regularized boundary conditions and bounce-back schemes in the context of LBM-RANS simulations with wall functions. The RANS framework is chosen for its relative simplicity and robustness, which facilitates controlled verification of numerical results. In contrast, WMLES approaches introduce multiple interacting numerical and physical factors, complicating the effort of isolating the effects of the boundary condition. To conduct this investigation, two canonical benchmark turbulent flows are selected: the fully developed turbulent channel flow and the zero-pressure-gradient turbulent boundary layer over a flat plate. These test cases provide a well-understood basis for evaluating the accuracy and robustness of wall boundary treatments in the context of turbulent modeling using the LBM approach. More complex test cases involving curved boundaries will be addressed in future work.

The paper is organized as follows. Section 2 introduces the Lattice Boltzmann method and describes with the turbulence and wall function models used in this work. Section 3 details the wall boundary condition schemes with wall functions. In Section 4, the performance of the different schemes is evaluated on two benchmark test cases, focusing on skin friction accuracy and turbulent velocity profile prediction. The effects of grid resolution and Reynolds number on both regularized and bounce-back approaches are also analyzed. Conclusions are presented in Section 5.

# 2. Numerical approach

## 2.1 Lattice Boltzmann Method

The LBM is based on a simplified kinetic model consisting in the discretization of the Boltzmann equation in space, time and velocity space and the simplification of the collision operator using, for example, the Bhatnagar-Gross-Krook (BGK) approximation and variations thereof. The LBM works with particle populations $|f_i\rangle = [f_1,..,f_Q]^T$, which are the probability distribution functions of particles with discrete velocity at a spatial location $\vec{x}$ and time *t*. The discrete velocities are defined on a regular lattice $\vec{c}_i = [|c_{ix}\rangle, |c_{iy}\rangle]$ (we focus on 2D for simplicity), and the evolution of the populations is governed by the Lattice-Boltzmann equation (LBE), which in dimensionless lattice units reads

$$f_i(\vec{x}+\vec{c}_i, t+1) = f_i(\vec{x}, t) + \Omega(f_i, f_i^{eq}, F_i) \quad (1)$$

where $\Omega$ is the collision operator and $|F_i\rangle$ are mesoscopic force distributions that account for external macroscopic body forces $\vec{F} = (F_\alpha) = [F_x, F_y]$. The LB equation (1) is usually solved in two sequential steps, i.e. the collision step, which shuffles the distribution functions at each node

$$f_i^*(\vec{x}, t) = f_i(\vec{x}, t) + \Omega(f_i, f_i^{eq}, F_i) \quad (2)$$

and the streaming step, which propagates the distribution functions to neighboring nodes according to their velocities,

$$f_i(\vec{x}+\vec{c}_i, t+1) = f_i^*(\vec{x}, t) \quad (3)$$

The (macroscopic) hydrodynamic variables density, $\rho$, flow velocity, $\vec{u} = (u_\alpha) = [u_x, u_y]$ and momentum flux tensor are obtained from discrete velocity moments of the populations as

$$\rho = \sum_i f_i \quad (4)$$

$$\rho u_\alpha = \sum f_i c_{i\alpha} + \frac{F_\alpha}{2} \quad (5)$$

$$\Pi_{\alpha\beta} = \sum_i f_i c_{i\alpha} c_{i\beta} \quad (6)$$

where $\alpha, \beta \in \{x, y\}$.

In this work, the universal central moment-based LBM of de Rosis et al. [36] is used. In this approach, the collision is performed in the space of central moments with a collision matrix, which, for the two-dimensional, nine-velocity (D2Q9) model takes the form

$$\mathbf{\Lambda} = diag[1,1,1,1,\omega,\omega,1,1,1] \quad (7)$$

where $\omega$ is a relaxation frequency.

The Chapman-Enskog expansion shows that the macroscopic equations solved by LBM in the hydrodynamic limit are the weakly compressible Navier-Stokes equations

$$\begin{aligned}&\partial_t\rho+\partial_\alpha(\rho u_\alpha)=0\\&\partial_t(\rho u_\alpha)+\partial_\gamma(\rho u_\alpha u_\gamma)=-\partial_\alpha p+\partial_\gamma(\rho\nu(\partial_\alpha u_\gamma+\partial_\gamma u_\alpha))+F_\alpha\end{aligned} \tag{8}$$

where the pressure and viscosity are given by

$$p=\rho c_s^2 \tag{9}$$

and

$$\nu=c_s^2\left(\omega^{-1}-\frac{1}{2}\right) \tag{10}$$

where $c_s$ is the speed of sound, which is $c_s=1/\sqrt{3}$ in lattice units for the D2Q9 velocity model. (In eq. (8) and in the remainder of the paper, summation over repeated dummy indices is understood). To compute turbulent flows with RANS-based models, the relaxation factor $\omega$ is modified in such a way that the viscosity in eq. (10) is replaced by the effective viscosity $\nu_{eff}=\nu+\nu_t$, where $\nu_t$ is the turbulent viscosity determined by the turbulence model.

## 2.2 Turbulence modeling

Turbulence modelling is addressed with the Spalart-Allmaras (SA) turbulence model [37] in the version described in [38], which has been selected in order to facilitate comparison with reference data in NASA's Turbulence Modeling Resource website [39].

In the SA model, the eddy viscosity $\nu_t$ is given in terms of a viscosity-like auxiliary variable $\tilde{\nu}$ (the SA working variable) as

$$\nu_t=\tilde{\nu}f_{v1},\quad f_{v1}=\frac{\chi^3}{\chi^3+c_{v1}^3},\quad \chi=\frac{\tilde{\nu}}{\nu} \tag{11}$$

where $\nu$ is the kinematic viscosity. $\tilde{\nu}$ obeys the following transport equation

$$\partial_t\tilde{\nu}+u_\alpha\partial_\alpha\tilde{\nu}=P-D+\frac{1}{\sigma}\left[\partial_\gamma((\nu+\tilde{\nu})\partial_\gamma\tilde{\nu}+c_{b2}\partial_\gamma\tilde{\nu}\partial_\gamma\tilde{\nu}\right] \tag{12}$$

where $\alpha,\gamma$ are Cartesian indices (and summation over repeated indices is understood), $u_\alpha$ are the Cartesian velocity components and

$$P=c_{b1}\tilde{S}\tilde{\nu},\quad D=c_{w1}f_w\left(\frac{\tilde{\nu}}{d}\right)^2 \tag{13}$$

are production and wall destruction terms. In eq. (13), $\tilde{S}$ is a modified vorticity

$$\tilde{S}=S+\frac{\tilde{\nu}}{\kappa^2d^2}f_{v2},\quad f_{v2}=1-\frac{\chi}{1+\chi f_{v1}} \tag{14}$$

where $S = \|\Omega\|$ is the magnitude of the vorticity, $d$ is the distance to the closest wall and the function $f_w$ is

$$f_w = g\left[\frac{1+c_{w3}^6}{g^6+c_{w3}^6}\right]^{1/6}, \quad g = r + c_{w2}(r^6 - r), \quad r = \min\left(\frac{\tilde{\nu}}{\tilde{S}\kappa^2 d^2}, r_{\lim}\right) \tag{15}$$

Finally, the constants of the model are

$$\begin{aligned} &c_{b1} = 0.1355, \quad c_{b2} = 0.622, \quad \sigma = 2/3, \quad \kappa = 0.41 \\ &c_{w1} = c_{b1}/\kappa^2 + (1+c_{b2})/\sigma, \quad c_{w2} = 0.3, \quad c_{w3} = 2.0 \\ &c_{v1} = 7.1, \quad r_{\lim} = 10.0 \end{aligned}$$

The transport equation (12) is discretized with finite differences. The convective term is discretized using a first-order upwind scheme, while the diffusive terms are discretized using standard second-order finite difference schemes. The vorticity in the production term (13) is computed from the velocity field of the LBM solution using second-order accurate finite differences. While the discretization is, a priori first-order accurate only, it is important to note that for the test cases considered (namely channel and flat plate flows), the convective terms have a negligible impact on the overall solution, as a result of which the scheme is second-order accurate in practice. The wall boundary condition is prescribed explicitly by the Dirichlet condition $\tilde{\nu}_B = \kappa \nu y_B^+$. The remaining boundary conditions are case-dependent, e.g. periodic conditions for the inlet/outlet boundaries in the case of turbulent channel flow.

In the near wall region, the model-consistent wall function described in [38] is used, which yields

$$\begin{aligned} u_{WF}^+(y^+) = \bar{B} &+ c_1 \log\left((y^+ + a_1)^2 + b_1^2\right) - c_2 \log\left((y^+ + a_2)^2 + b_2^2\right) \\ &- c_3 \arctan\left(b_1/(y^+ + a_1)\right) - c_4 \arctan\left(b_2/(y^+ + a_2)\right) \end{aligned} \tag{16}$$

for the velocity, where

$$\begin{aligned} &\bar{B} = 5.033908790505579 \\ &a_1 = 8.148221580024245, \quad a_2 = -6.9287093849022945 \\ &b_1 = 7.4600876082527945, \quad b_2 = 7.468145790401841 \\ &c_1 = 2.5496773539754747, \quad c_2 = 1.3301651588535228 \\ &c_3 = 3.599459109332379, \quad c_4 = 3.6397531868684494 \end{aligned}$$

and

$$\frac{du_{WF}}{dy} = \frac{u_\tau^2}{\nu}\frac{du_{WF}^+}{dy^+} = \frac{u_\tau^2}{\nu}\left(\frac{c_{v1}^3 + (\kappa y^+)^3}{c_{v1}^3 + (1+\kappa y^+)(\kappa y^+)^3}\right) \tag{17}$$

for the velocity gradient, where $u_\tau$ is the friction velocity at the wall.

## 3 Wall boundary conditions for LBM-RANS with wall functions

At boundary nodes, the streaming step involves populations that originate outside the domain (populations corresponding to grid directions 1, 2 and 8 in the D2Q9 example of Figure 1). These missing populations must be provided to fully define the LBM scheme. To this end, they have to be tailored to incorporate information of the physical flow boundary condition through the computed macroscopic quantities. In the case of solid walls, two different types of boundary conditions are usually found in the literature: regularized boundary schemes, in which all the populations are reconstructed using Chapman-Enskog expansions, and bounce-back approaches, in which populations are reconstructed by reflection.

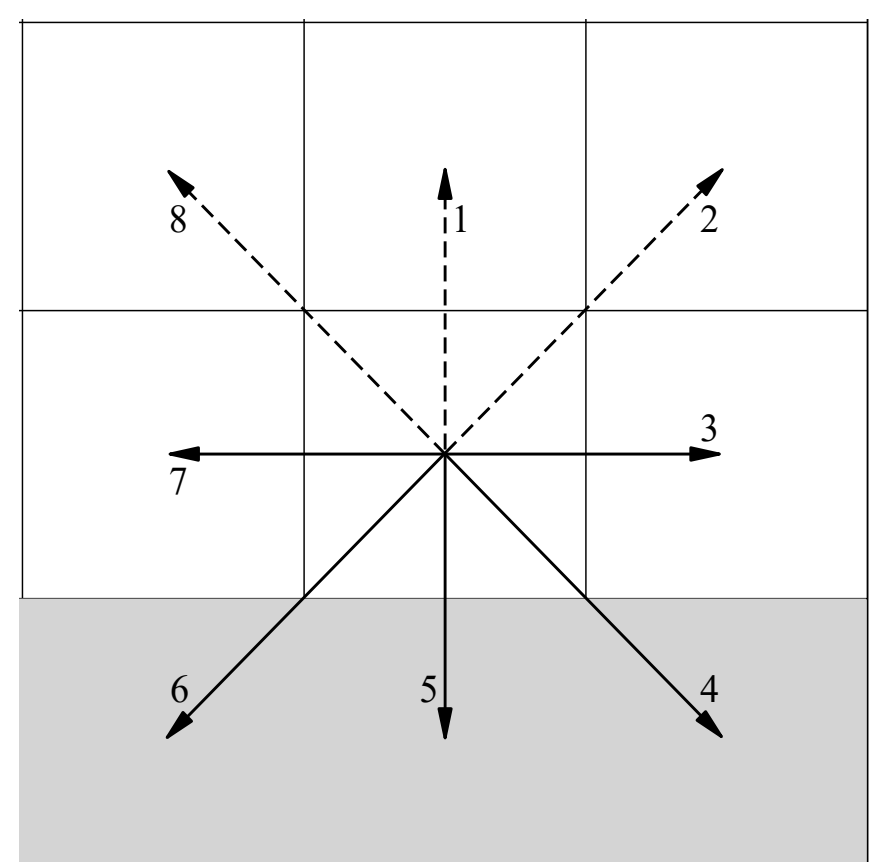

Figure 1 Sketch of a boundary node for the 2D LBM with a D2Q9 lattice. Dashed arrows represent populations streamed from outside the domain

The regularized boundary scheme is a Dirichlet-type boundary condition. In its most common implementation, all distribution functions at boundary nodes are reconstructed (or regularized) based on their equilibrium and non-equilibrium components. Through this regularization process, the boundary populations are implicitly constrained by the macroscopic flow variables: density and velocity and the strain rate tensor, which is related to boundary velocity gradients.

The bounce-back method, on the other hand, is a link-wise approach based on the bounce-back scheme. In this scheme, only the distribution functions reflected from the wall are replaced. For turbulent flows modeled with LBM coupled to a RANS closure and wall functions, the bounce-back formulation provides a subtle but effective mechanism to impose a turbulent diffusion equation at the boundary. This is achieved by prescribing an appropriate slip velocity at the wall, consistent with the assumptions of the wall function model.

Both approaches differ fundamentally in their formulation and physical interpretation. Performing a rigorous theoretical analysis of their respective accuracy is highly complex, even for simple laminar flows (see for example [40]), and lies beyond the scope of this paper. Instead, the objective of the present study is to provide practical insight into their performance through carefully designed numerical experiments on two well-controlled benchmark cases.

### 3.1 Regularized wall boundary condition

The regularized boundary condition for flat surfaces was originally proposed by Latt et al. in [16]. For full details, the reader is referred to the original publication. Here, we summarize only the main result. The regularized approach is based on a multi-scale Chapman-Enskog (CE) expansion analysis of the lattice Boltzmann equation (LBE) model [23]. This expansion assumes that the population distributions can be expressed as a perturbation series $f = f_i^{(0)} + f_i^{(1)} + f_i^{(2)} + \dots$, where the zeroth-order term corresponds to the equilibrium distribution and the higher-order terms scale with successive powers of the Knudsen number (Kn), i.e. $f_i^{(n)} \sim (Kn)^n$, with $Kn \ll 1$. It has been shown that truncating the expansion after the first two terms and applying multi-scale analysis to the LBE leads to an asymptotic recovery of the Navier-Stokes equations. Furthermore, the CE expansion establishes a connection between the particle distribution functions and the macroscopic hydrodynamic variables for isothermal flows, expressed as:

$$f_i^{(0)} = f_i^{eq}(\rho, \vec{u}) \tag{18}$$

and

$$f_i^{(1)} \approx -\frac{w_i \rho}{\omega c_s^2} Q_{i,\alpha\gamma} S_{\alpha\gamma} \tag{19}$$

where $w_i$ are the lattice weights, $S_{\alpha\gamma} = \frac{1}{2}\left(\partial_\alpha u_\gamma + \partial_\gamma u_\alpha\right)$ is the strain tensor and $Q_{i,\alpha\gamma} = c_{i\alpha} c_{i\gamma} - c_s^2 \delta_{\alpha\gamma}$.

Thus, regularization in this context serves as a Dirichlet-type boundary condition, in which all the distribution functions at a boundary node $\vec{x}_B$ are replaced by reconstructed ones based on the expressions (18) and (19)

$$f_i(\vec{x}_B, t+1) = f_i^{eq}(\rho_B, \vec{u}_B) + f_i^{(1)}(\rho_B, \nabla \vec{u}_B) \tag{20}$$

A key ingredient of the regularized boundary condition given by equation (20), particularly in the context of turbulent flows with wall functions, is the accurate evaluation of the term $f_i^{(1)}$. Precise computation of the velocity gradients involved in the strain rate tensor $S_{\alpha\gamma}$ is crucial, since the boundary point typically lies within a region characterized by steep velocity gradients. This challenge will be addressed in detail in Section 3.1.1.

In the following, we spell out in detail the steps required to implement the wall turbulent boundary condition on flat boundaries for the LBM with cell-centered approach used in this study

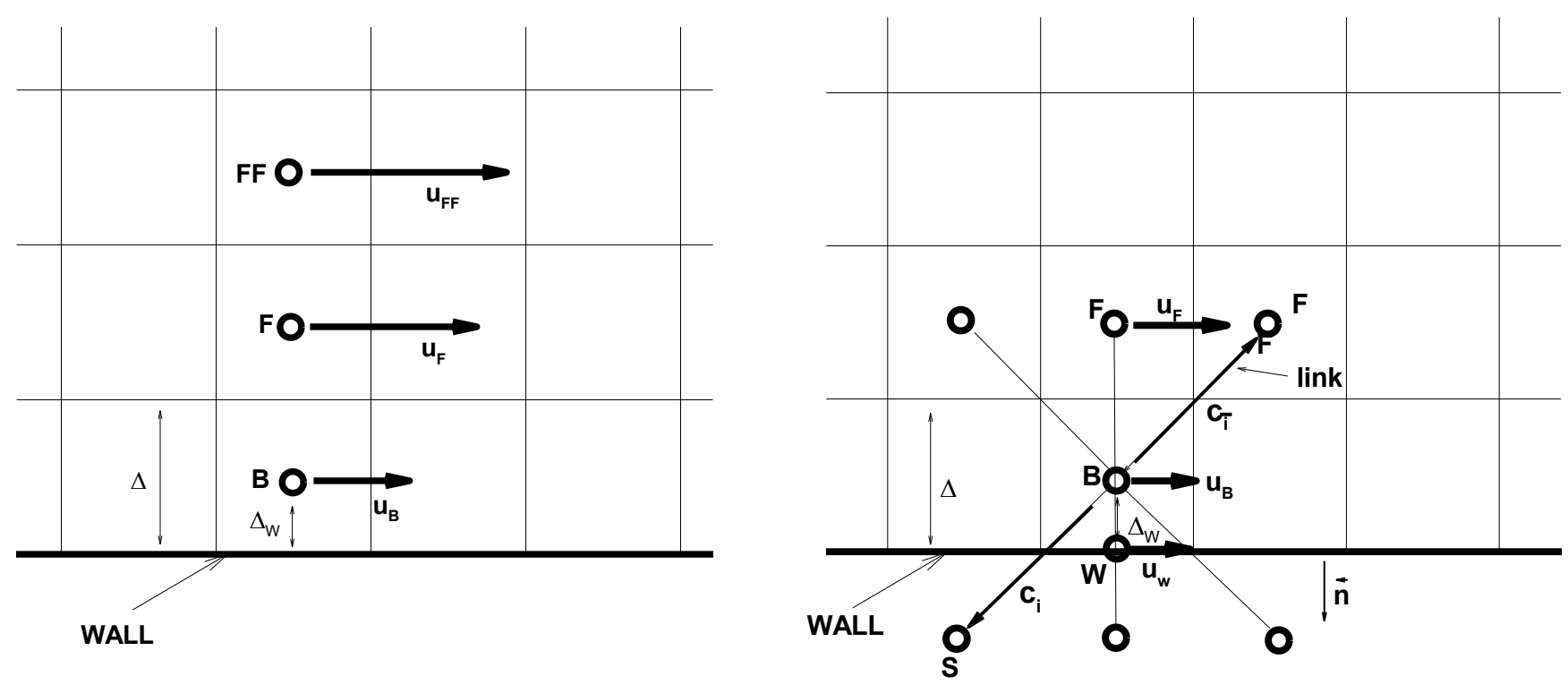

(a) (b)

Figure 2 (a) Regularized-type boundary condition: Location of nodes W, B, F and FF, i.e. wall, boundary and next-to-boundary nodes along the wall normal direction. (b) Bounce-back type condition: example of nodes in a diagonal link.

The computation of the equilibrium part of the reconstructed population at node B (first interior node off the boundary, see Figure 2(a)) has become standard lore in the literature (see e.g. [18]), but we repeat it here for completeness. After the streaming step, the density and velocity at a reference point (which in this work is taken to be the second point off the wall, node F) are computed taking the corresponding moments of the populations,

$$\rho_F = \sum_i^Q f_i(\vec{x}_F, t+1) \tag{21}$$

and

$$\left(\rho\vec{u}\right)_F = \sum_i^Q f_i(\vec{x}_F, t+1)\cdot\vec{c}_i + \frac{\vec{F}}{2} \tag{22}$$

The tangential component of the velocity is trivially computed for the test cases considered in this work (flat surfaces) as

$$\left(u_t\right)_F = \vec{u}_F \cdot \vec{t}_F \tag{23}$$

where $\vec{t}_F = (1,0)$. Once the tangential velocity is known at the reference point, the friction velocity can be computed using a Newton-Raphson method to solve to solve for $u_\tau$ in the relation $u^+_{WF}(y^+_F)$ given by the law of the wall (16). Next, the tangential component of the velocity at the boundary node is computed using again the analytical wall function

$$\vec{u}_B = u_\tau u^+_{WF}(y^+_B)\vec{t}_F \quad (24)$$

with the current value of $u_\tau$. To reconstruct the equilibrium population, the density at node B is also required. A zero order extrapolation from node F is typically used in curved geometries, but here we use, following [18]

$$\rho_B = \frac{1}{1+\vec{u}_B\cdot\vec{n}}(f_\parallel + 2f_\perp) \quad (25)$$

in order to conserve mass strictly. In eq. (25), $\vec{n}$ is the outward normal unit vector at the boundary,

$$f_\parallel = \sum_{i|\vec{c}_i\cdot\vec{n}=0} f_i \quad (26)$$

and

$$f_\perp = \sum_{i|\vec{c}_i\cdot\vec{n}>0} f_i \quad (27)$$

Once the density and velocity at node B are known, the equilibrium population is computed using the corresponding expression from the LBM model [36], while the non-equilibrium part of the populations at node B are computed as

$$f_i^{neq}(\vec{x}_B, t+1) \approx f_i^{(1)}(\rho_B, \nabla\vec{u}_B) = -\frac{w_i\rho_B\tau_{eff}}{c_s^2}Q_{i,\alpha\gamma}S_{\alpha\gamma} \quad (28)$$

where the effective relaxation time is given by $\tau_{eff} = 1/2 + \nu_{eff}/c_s^2$, where $\nu_{eff} = \nu + \nu_t(\vec{x}_B, t)$ and $\nu_t(\vec{x}_B, t)$ is the turbulent eddy viscosity computed by the RANS turbulence model at the previous step. Since we use the macroscopic information given by the strain tensor $S_{\alpha\gamma} = \left(\partial_\alpha u_\gamma + \partial_\gamma u_\alpha\right)/2$ in the non-equilibrium population reconstruction, the gradients of velocities must be estimated[1]. In this paper, we use for the *x*-direction (running parallel to the wall boundary) second order symmetric finite differences. The velocity gradient in the normal wall direction needs a special treatment that will be addressed in the next section. Finally, all the populations at the boundary node are replaced by the reconstructed populations using (20).

[1] The strain tensor can be obtained directly from mesoscopic information but we will not follow that route here.

### 3.1.1 Wall normal velocity gradient reconstruction methods

The accurate reconstruction of the velocity gradients used to estimate the strain rate tensor in the non-equilibrium distribution function eq. (19) is a critical factor influencing the accuracy of the regularized boundary condition, as will be shown later. In turbulent boundary layers, the wall-normal component of the velocity gradient typically dominates. Therefore, the method used to evaluate this component has a significant impact on the overall performance of the boundary condition scheme.

Several strategies can be considered for reconstructing the wall-normal velocity gradient. Given that we are operating within the wall function framework, where an analytical expression for the velocity profile is available, one intuitive approach is to directly use the corresponding analytical derivative to estimate the strain rate. However, it will be shown that this method is not suitable for regularized boundary schemes, as it introduces inaccuracies. An alternative approach is to compute the wall-normal derivative using a finite difference scheme based solely on the local flow field, without incorporating any explicit information from the wall function. This method is consistent with the original formulation of regularized schemes but also leads to poor results when applied to RANS simulations with wall functions. Motivated by these limitations, we investigate a hybrid strategy that leverages both sources of information, the numerical flow field and the analytical wall function profile. To this end, we introduce a blended reconstruction of the wall-normal velocity gradient, defined by the following expression:

$$\partial_y u\big|_B = \underbrace{\beta\,\partial_y u_{WF}(y)\big|_B}_{\substack{\text{analytical derivative}\\ \text{from wall function}}} + \underbrace{(1-\beta)\,\delta_y^{(n)} u\big|_B}_{\substack{\text{numerical derivative}\\ \text{from the flow}}} \tag{29}$$

where $u$ represents the wall-parallel velocity (and we will stick to that notation throughout the remainder of this paper for simplicity). In eq. (29), the analytical derivative $\partial_y u_{WF}$ is computed directly from eq. (17), while $\delta_y^{(n)} u$ stands for the wall-normal derivative estimated using an *(n)-order* finite difference scheme applied to the computed velocity field obtained in the previous iteration. Preliminary tests have shown that a first-order accurate scheme leads to insufficient accuracy and is therefore not recommended. In this paper, we use a one-sided second-order accurate scheme

$$\delta_y^{(2)} u\big|_B = \frac{1}{2\Delta}\left(-3u_B + 4u_F - u_{FF}\right) \tag{30}$$

where $\Delta$ is the boundary cell size (see Figure 2(a)). Note that a third fluid node FF is needed to evaluate the second order formula (30). Finally, $\beta$ is a blending parameter whose value will be calibrated by comparison against a reference solution, as described in Section 4.1.2.1. For the sake of brevity, we refer to this approach as the Hybrid Regularized Finite Difference reconstruction (HRFD) method in the remainder of the paper. Eq. (29) enables the examination of alternative strategies for reconstructing the wall-normal velocity gradient. For instance, by setting $\beta = 0$ we recover the original regularized boundary condition formulation based solely on finite difference approximations of the velocity field [16], which was initially applied to laminar flows. This setting also corresponds to simplified variants of regularized schemes presented by Wilhelm et al. [19] and Feng et al. [17], although these authors employed a one-sided, first-order finite difference operator rather than the second-order scheme adopted in the present study.

#### 3.1.2 The Cai-Degrigny velocity gradient reconstruction

Cai et al. [20] and Degrigny et al. [21] proposed an alternative methodology for estimating the wall-normal velocity gradient in the context of regularized boundary condition schemes. Their work, conducted within the more complex framework of immersed boundary methods on Cartesian grids, is particularly valuable as it provides one of the few detailed implementations of LBM wall modeling within a RANS framework. As highlighted in [21], exclusive use of the analytical derivative of the wall function can compromise the overall accuracy of the numerical solution. To address this issue, the authors compute the wall-normal velocity gradient with third-order finite-differences using wall function data, i.e.,

$$\partial_y u_{WF}\big|_B \approx \delta_y^{(3)} u_{WF}\big|_B = \frac{u_\tau}{\Delta}\left(-\frac{11}{6}u_{WF}^+(y_B^+) + 3u_{WF}^+(y_B^+ + \Delta^+) - \frac{3}{2}u_{WF}^+(y_B^+ + 2\Delta^+) + \frac{1}{3}u_{WF}^+(y_B^+ + 3\Delta^+)\right) \tag{31}$$

We will refer to this approach as the *Cai-Degrigny method* throughout the rest of the paper. It is worth noting that this method can be regarded as a particular case of the HRFD scheme by setting the blending parameter $\beta = 1.0$ and replacing the analytical derivative with a third-order finite difference operator, i.e.,

$$\underbrace{\partial_y u_{WF}\big|_B}_{analytical\ derivate} \approx \underbrace{\delta_y^{(3)} u_{WF}\big|_B}_{numerical\ derivate\ of\ the\ wall\ function} \tag{32}$$

This method will be assessed in the comparative analysis between regularized based boundary schemes that will be presented in Section 4.1.2. A summary of the regularized schemes evaluated in this work is shown in Table 1.

### 3.2 Turbulent wall slip velocity bounce-back-based boundary condition

In order to cover the different alternatives of LBM wall boundary condition formulations for turbulent flows, a bounce-back formulation will also be considered. The proposed bounce-back scheme is based on Nishimura et al. [28] adapted to work together with RANS turbulence models and wall functions.

The classical formulation of the half-way bounce-back rule assumes that the wall lies halfway between the fluid boundary node and a virtual solid node. This model is often explained using the intuitive picture of mesoscopic populations (representing particle distributions) traveling along a lattice link toward the wall with velocity $\vec{c}_i$, reflecting at the wall, and returning along the opposite direction with velocity $\vec{c}_{\hat{i}} = -\vec{c}_i$ , thereby conserving momentum. This amounts to replacing the standard streaming step for the missing populations by

$$f_{\hat{i}}(\vec{x}_B, t+1) = f_i^*(\vec{x}_B, t) \qquad \vec{c}_i \cdot \vec{n} > 0 \tag{33}$$

where $f_i^*$ denotes the post-collision population (see Figure 3b). Eq. (33) can be generalized to moving walls or slip velocities [15, 23, 25, 26] with a correction to account for the momentum exchange with the moving wall so that the outcome respects Galilean invariance. The population at the boundary node after bounce-back is computed as:

$$f_{\hat{i}}(\vec{x}_B, t+1) = f_i^*(\vec{x}_B, t) - 2 w_i \rho_w \frac{\vec{c}_i \cdot \vec{u}_w}{c_s^2} \qquad \vec{c}_i \cdot \vec{n} > 0 \tag{34}$$

where $\vec{u}_w$ is the relative velocity between the wall and the fluid, $\rho_w$ is the density at the wall and the coefficients $w_i$ (known as weights) depend on the set of velocities of the LB model (see Figure 3c). One way to prove eq. (34) is to transform to the rest frame of the wall, perform bounce-back there, and transform back to the initial frame. Notice that eq. (34) reduces to eq. (33) for a stationary boundary.

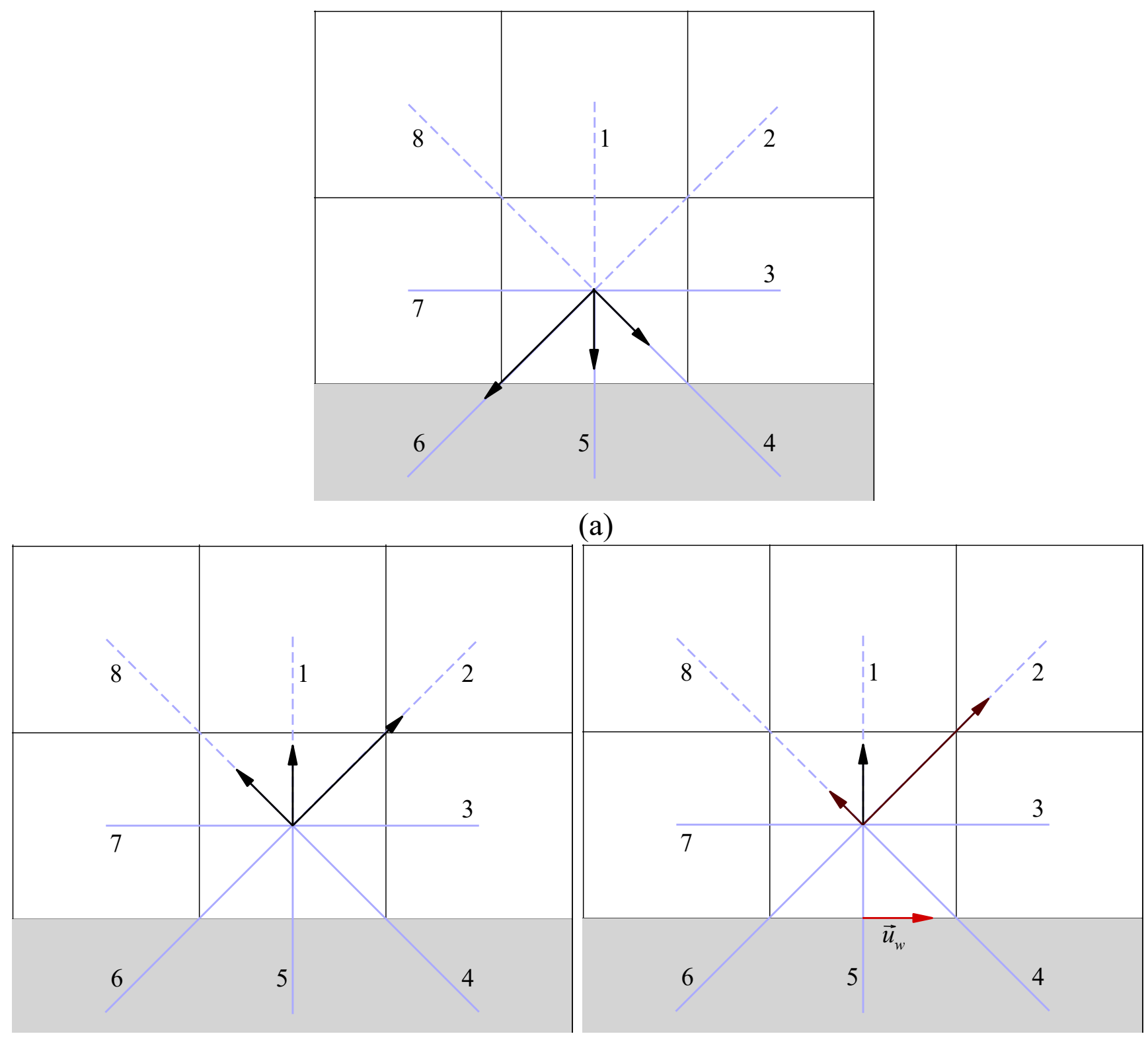

(b) (c)

Figure 3 Scheme of the bounce-back rule. Violet lines represent lattice velocity directions (dashed lines correspond to unknown populations). Black arrows represent populations. (a) Populations 4, 5 and 6 before bounce-back (b) Populations 1, 2, and 8 after bounce-back with zero slip velocity. (c) Same as (b) with non-zero slip velocity.

#### 3.2.1 Derivation of the slip velocity bounce-back boundary condition

The idea of the slip velocity bounce-back boundary condition is to assume a fictitious slip velocity in eq. (34) whose value is fixed to enforce the appropriate turbulent equations. In the case of a turbulent boundary layer with wall functions, the governing equation obeyed at the fluid boundary node is the turbulent diffusion equation:

$$\partial_y((\nu+\nu_t)\partial_y u)\big|_B = 0 \tag{35}$$

Eq. (35) is a statement of constant total shear stress and can be integrated to yield

$$(\nu+\nu_t)\partial_y u\big|_B = u_\tau^2 \tag{36}$$

which is the desired boundary condition at the wall. In order to impose eq. (36), we investigate the effect of the boundary condition eq. (34) on the fluid velocity. Rigorously explaining the macroscopic behavior of the bounce-back rule (as any other mesoscopic model) requires techniques such as the Chapman-Enskog analysis [23] [40], which lie beyond the scope of this work. However, it can be shown that, under simplifying assumptions (namely stationary, unidirectional flow, BGK collision operator and a linearized equilibrium distribution function), a second-order Chapman–Enskog expansion of the half-way bounce-back rule (34) implicitly yields the following kinematic (macroscopic) boundary condition:

$$u_w = u_B - \frac{\Delta}{2}\partial_y u\Big|_B + \frac{2}{3}\Delta^2 E(\tau)\partial_y^2 u\Big|_B \tag{37}$$

where $E(\tau) \approx (\tau - 1/2)^2$ and $\tau = \omega^{-1} = \overline{\tau}/\Delta t$ is the dimensionless relaxation time which is related to the viscosity by eq. (10). A derivation of eq. (37), following [23], is summarized in the appendix. The third term in eq. (37) is known in the specialized literature as the numerical slip velocity and may have a significant effect on the accuracy of the boundary condition for laminar flows at low and moderated Reynolds numbers [40]. In principle, the effect of this term can be eliminated by using specifically tuned collision operators (TRT, MRT, etc.) but we will not follow that path here. Instead, we assume that this term can be neglected, an assumption that we will justify momentarily, so that eq. (37) yields for the velocity gradient the following result

$$\partial_y u\Big|_B \approx (u_B - u_w)/\Delta_w \tag{38}$$

where $\Delta_w = \Delta/2$. We recall that this is the macroscopic boundary condition realized by the bounce-back rule (34). Using eq. (38) in eq. (36) to solve for the slip wall velocity yields

$$u_w = u_B - \frac{u_\tau^2}{\nu + \nu_t\big|_B}\Delta_w \tag{39}$$

where $\nu_t\big|_B = \kappa\nu y_B^+$ from the turbulence model boundary condition and $u_\tau$ is obtained from the analytical wall function via a Newton-Raphson algorithm as in the regularized approach described in 3.1. Eq. (39) is the value of the (wall-parallel) slip velocity required to enforce the turbulent diffusion equation (36) at the boundary node.

**3.2.2 Remarks on the validity of the derivation**

Even though the Chapman–Enskog expansion (37) is not strictly applicable to our case, since the collision model and equilibrium functions used in this work differ from the BGK model, it is still instructive to examine the order of magnitude of the coefficient $E(\tau)$ in the third term in the expansion. This analysis helps to explain some of the results presented in Section 4 and serves to justify why it can be neglected when deriving eq. (39). In the simulations performed in this paper, the relaxation time is determined from the input parameters as

$$\tau = 1/2 + \sqrt{3}M / (\mathrm{Re}_L\, Kn) \tag{40}$$

where $M$ is the free-stream Mach number, $\mathrm{Re}_L = UL/\nu$ is the Reynolds number relative to a characteristic scale of the flow $L$, $Kn = \Delta / L$ is the numerical Knudsen number and an acoustic scaling has been assumed. The turbulent flows considered in this paper have Reynolds numbers in the order of $O(10^5 - 10^7)$. Since we are using wall functions, the corresponding spatial resolutions lead to numerical Knudsen numbers in the range $O(10^{-2} - 10^{-4})$. Therefore, $E(\tau) \approx O(M^2 / (\mathrm{Re}_L\, Kn)^2) \approx O(10^{-8}) \ll 1$. Consequently, the third term in eq. (37) is negligible in comparison to the slip velocity $u_w$ so that eq. (38) holds.

### 3.2.3 Interpolated bounce-back conditions

In order to consider grid configurations where the boundary node is not exactly half-way between grid points, we consider two linearly interpolated bounce-back methods. The interpolated bounce-back scheme from Bouzidi et al [25] is used to obtain the results of section 4.1.2.5, where the sensitivity to wall location is examined. In Bouzidi's method the reflected incoming populations are obtained as

$$f_{\bar{i}}(x_B, t+1) = \begin{cases} \dfrac{1}{2q_i} f_i^*(x_B, t) + \left(\dfrac{2q_i - 1}{2q_i}\right) f_{\bar{i}}(x_F, t+1) - \dfrac{1}{q_i} \rho_w w_i \dfrac{\vec{c}_i \cdot \vec{u}_w}{c_s^2} & q_i \geq \dfrac{1}{2} \\ 2q_i f_i^*(x_B, t) + (1 - 2q_i) f_i^*(x_F, t) - 2\rho_w w_i \dfrac{\vec{c}_i \cdot \vec{u}_w}{c_s^2} & q_i < \dfrac{1}{2} \end{cases} \tag{41}$$

where $f_i^*$ are post-collision distributions and $q_i = |\vec{x}_B - \vec{x}_W| / |\vec{x}_B - \vec{x}_S|$. When $q_i = 0.5$ the halfway bounce-back is recovered. In this work we assume the approximation $\rho_w \approx \rho_B$

On the other hand, for the flat plate test case in Section 4.2, a different interpolated bounce-back scheme was employed. Bouzidi's scheme was tried first but it led to convergence issues, primarily due to pronounced odd-even oscillations near the leading edge of the flat plate, where symmetry and wall boundary conditions intersect. To address these numerical instabilities, we adopted the interpolated bounce-back method proposed by Yu et al. [26]. Yu's linear interpolation approach uses a single expression for all $q_i$ which is given by

$$\begin{aligned} f_{\bar{i}}(x_W, t+1) &= f_i(x_F, t+1) + (1 - q_i) f_i^*(x_B, t) - 2\rho_w w_i \frac{\vec{c}_i \cdot \vec{u}_w}{c_s^2} \\ f_{\bar{i}}(x_B, t+1) &= f_{\bar{i}}(x_W, t+1) + \left(\frac{q_i}{q_i + 1}\right)\left(f_{\bar{i}}(x_F, t+1) - f_{\bar{i}}(x_W, t+1)\right) \end{aligned} \tag{42}$$

where $W$ denotes the wall node (see Figure 1(b)). We refer to [26] for a detailed derivation of the approach. It is important to note that, unlike Bouzidi's scheme, this method does not recover the half-way bounce back expression when $q_i = 0.5$ but retains the distribution function at node *F*, $f_i(x_F, t+1)$, in the reflected population, which is essential to suppress odd-even oscillations in the solution, particularly near boundary intersections.

### 3.3 Summary of boundary schemes

Table 1 summarizes the boundary schemes along with references to original sources where they were proposed and validated, while Figure 4 contains a visual roadmap of the different strategies involved.

Table 1. Summary of boundary conditions studied in this paper and related boundary conditions in the literature

| Boundary condition | Features & references |
|---|---|
| Regularized with FD evaluation of gradients | Finite difference regularized b.c (BC4 of paper [16]) laminar flow [19] [17] Laminar & LBM-RANS |
| HRFD with $\beta = 0.0$ | Present paper RANS based + wall function |
| Cai-Degrigny gradient reconstruction | Corrected wall normal gradient [21] [20] LBM-RANS + wall function |
| HRFD with $\beta = 1.0$ & third order FD of wall function | Present paper: LBM-RANS based + wall function |
| HRFD with calibrated $\beta$ | Present paper: LBM-RANS + wall function |
| Velocity slip bounce-back | [27] [28] LBM-Implicit -WMLES Present paper: LBM-RANS +wall function |

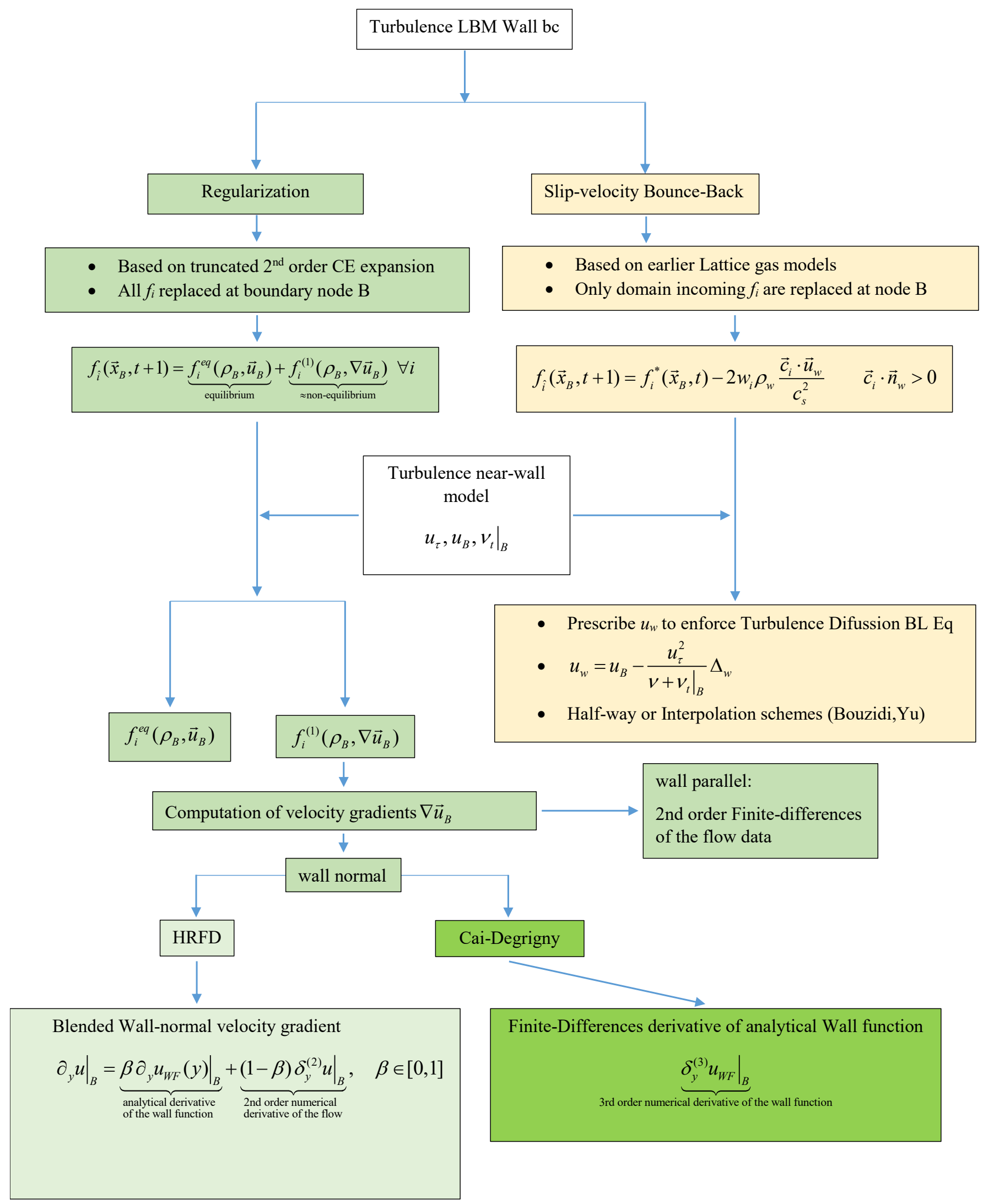

Figure 4. Flowchart of the implementation of the boundary schemes considered in this paper

## 4. Results and discussion

Two benchmark test cases have been selected to evaluate and compare the performance of the proposed wall boundary condition schemes. The first is the canonical periodic turbulent channel flow. This configuration is particularly well-suited for isolating the effects of the wall boundary treatment, as the remaining boundary condition, periodicity at the inlet and outlet, is trivially imposed [23] and does not influence the flow development. Due to the periodicity condition the flow must be driven by a constant external uniform force in the stream-wise direction. The magnitude of this forcing term corresponds to the pressure gradient required to sustain a fully developed turbulent channel flow at a specified friction Reynolds number $Re_\tau$. This pressure gradient is proportional to the square of the friction velocity normalized by the half-channel height [41]. During the simulation, the wall shear stress must evolve to balance the applied forcing, and the resulting computed friction velocity provides a direct quantitative measure of the accuracy of the wall boundary condition for a given turbulence model.

The second test case is the zero-pressure-gradient turbulent boundary layer (ZPG-TBL) over a flat plate, as proposed in NASA's Turbulence Modeling Resource (TMR) website [39]. This test case represents the simplest form of an external aerodynamic turbulent boundary layer flow, as it avoids geometric curvature and induced pressure gradients. As such, it provides a clean scenario for assessing the behavior and performance of wall boundary schemes in isolation, without requiring auxiliary modeling techniques such as flow interpolation. Moreover, the underlying assumptions of the Spalart-Allmaras wall function model (namely local equilibrium and negligible pressure gradient) are satisfied, making this case well-suited for validation purposes.

We finally note that we have also tested the same cases with a different turbulence model and wall function,

## 4.1 Turbulent channel flow

### 4.1.1 Test case description

We will compare the performance of the boundary conditions described in section 3 (see Table 1) on a turbulent channel flow between two parallel walls. The size of the domain is $5h$ in the streamwise ($x$) direction and $2h$ in the wall-normal ($y$) direction, where $h$ is the channel half-width. The friction Reynolds number $Re_\tau$ is an input for the computation. We also prescribe the input density and the dynamic viscosity, $\rho_0 = 1$ kg/m$^3$ and $\nu = 1.5\times10^{-5}$ m$^2$/s, respectively and we take $h = 1$ m. We impose periodic inlet/outlet conditions. The target $Re_\tau = (u_\tau h)/\nu$ is imposed by adding a constant external force along the streamwise direction to the LBE equation, i.e. $\mathbf{F} = \partial_x p = \rho_0 \mathbf{g} = \left(\rho_0 (u_\tau^2)_{target} / h\right)(1,0)^T$. The bulk velocity, which is defined as $U_b = (1/2h)\int_0^{2h} u(y)dy$, the bulk Reynolds number, $Re_b = U_b(2h)/\nu$, and the computed friction velocity $(u_\tau)_{comp}$, are byproducts of the simulation. The flow is initialized with the bulk velocity obtained from $Re_\tau$ using Dean's correlation $Re_b = (8.0/0.073)^{(4/7)} Re_\tau^{8/7}$ [42]. All LBM simulations are run with an acoustic time step (the time step is fixed in terms of a reference velocity, a prescribed –small– Mach number and the spatial grid resolution, $\Delta t = M\Delta x/(\sqrt{3}U_{ref})$. Numerical simulations are converged with the criterion $\Delta u_\tau^{n+1}/u_\tau^n < 10^{-4}$. Table 2 summarizes the numerical parameters used for the simulations in section 4.1.2. In order to perform a comparative analysis of the different boundary condition schemes, we will focus first on $Re_\tau = 4200$. The reference Direct Numerical Simulation (DNS) results at $Re_\tau \approx 4200$ for verification analysis are taken from the DNS database [43].

Table 2. Flow parameters covered in the simulations in this investigation. Values are obtained with Dean's correlation. (*) Reference data for comparison and verification are obtained from DNS results *[43]*.

| $Re_\tau$ | $u_\tau^*$ **[m/s]** | g**[m/s²]** | $Re_b$ | $U_b$ **[m/s]** | $C_{fb}$ | $y_{first}(y^+=50)$ |
|---|---|---|---|---|---|---|
| 2000 | 0.0300 | $9\times10^{-4}$ | $8.6734\times10^{4}$ | 0.6505 | 0.00425 | 0.0250 |
| 4200(*) | 0.0630 | 0.0040 | $2.0251\times10^{5}$ | 1.5188 | 0.00344 | 0.0119 |
| 8000 | 0.1200 | 0.0144 | $4.2292\times10^{5}$ | 3.1719 | 0.00286 | 0.0062 |
| 20000 | 0.3000 | 0.09 | $1.2052\times10^{6}$ | 9.0387 | 0.00203 | 0.0025 |

### 4.1.2 Results

#### 4.1.2.1 Effect of $\beta$ in the hybrid reconstruction HRFD approach

We wish to begin by analyzing the impact of the blending parameter on the hybrid reconstruction of the wall-normal velocity derivative defined in eq. (29). Simulations for the turbulent channel flow at $Re_\tau = 4200$ are conducted on a uniform grid with $N = 20$ points per half-channel, where the boundary node is positioned at $y_B = \Delta/2$ and $\Delta = 0.025h$, which is equivalent to $y_B^+ \approx 100$ in wall units. In practical applications, this corresponds to a typical medium-grid resolution in RANS simulations with wall functions. Figure 5(a) shows a comparison of the computed friction velocities for several values of the blending parameter ranging from 0 (indicating that the wall-normal velocity gradient is reconstructed purely using finite differences from the flow field) to 1, where the gradient is obtained entirely from the analytical derivative provided by the wall function. It can be observed that the computed friction velocity is quite sensitive to the method used to reconstruct the wall-normal gradient. Specifically, using finite differences from the flow solution tends to under-predict both the friction velocity and the velocity profile (see Figure 5(b)) compared to the reference solution. It will be shown later that this outcome is linked to the underestimation of the wall-normal velocity derivative at the boundary node. Conversely, using only the analytical derivative leads to a significant over-prediction of both the computed friction velocity and the corresponding velocity profile when compared to the reference DNS solution. Figure 5(c) displays semi-logarithmic plots of the velocity profiles, normalized by their respective computed friction velocities. Although the velocity at the boundary point satisfies the classical log-law, as expected from the wall-function approach, the velocity profiles in both the log-law and defect regions deviate from the expected behavior, particularly for extreme values of the blending parameter. Nonetheless, the results in Figure 5 also indicate that there exists an intermediate value of the blending parameter that yields good agreement with the reference solution. This includes the accurate prediction of the friction velocity, the velocity profile (b), and the semi-log velocity distribution (c), suggesting that the turbulent flow solution remains consistent with the external forcing defined by the input parameters.

This observation motivates the search for a calibrated value of the blending parameter that can reproduce as closely as possible the reference solution provided by DNS simulations. By performing a parametric sweep over the blending parameter in eq. (29), a specific value, $\beta^* = 0.28$ was determined for which the resulting computed friction velocity closely matches the prescribed target value (see Figure 6(a)). When employing this value, the resulting velocity profile exhibits strong agreement with the DNS reference results, both when normalized by the input friction velocity and by the computed one, both of which are effectively identical in this case (Figure 6 (b)-(c)). This agreement confirms the consistency of the hybrid reconstruction with the underlying wall-model assumptions. The calibrated value $\beta^*$ will therefore serve as the reference blending parameter in the HRFD methodology for the remainder of this study.

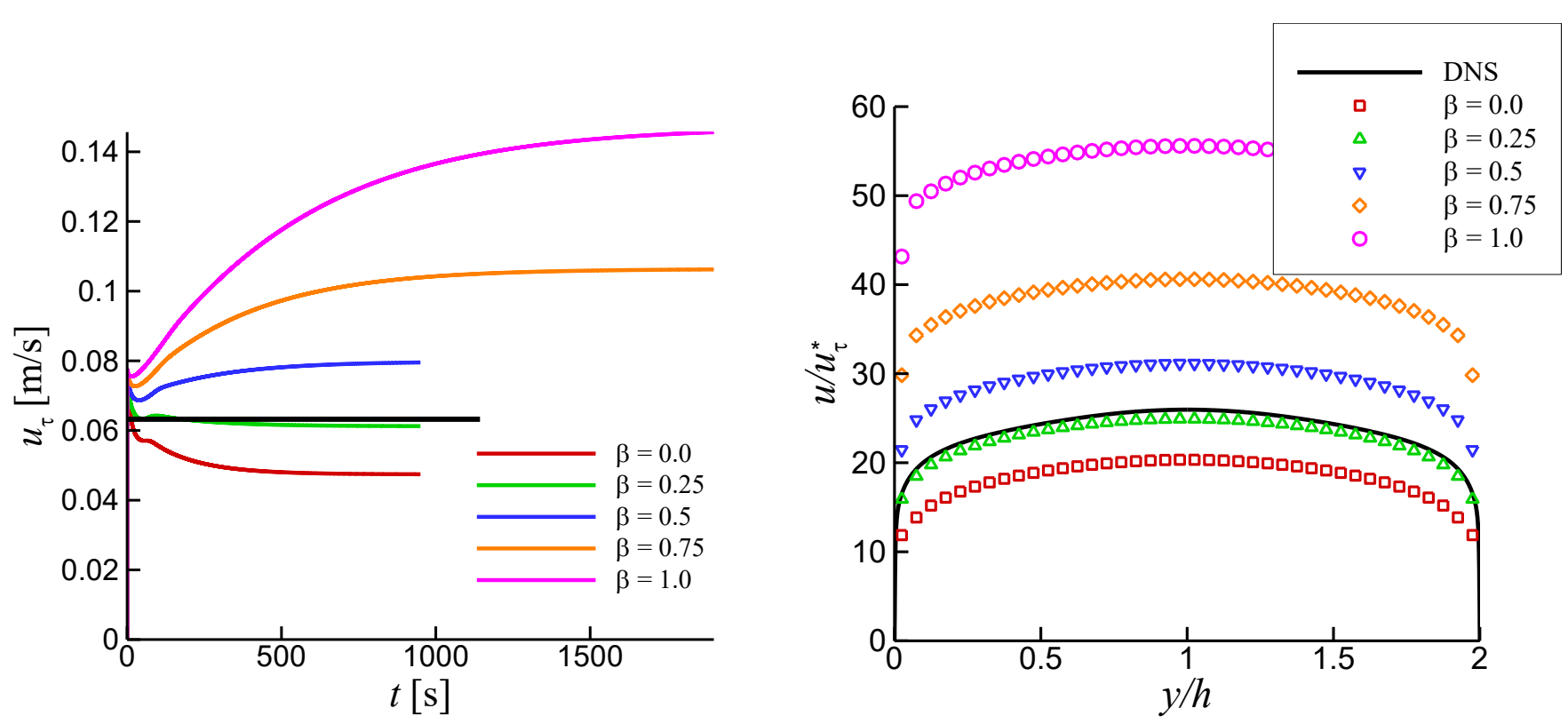

(a) (b)

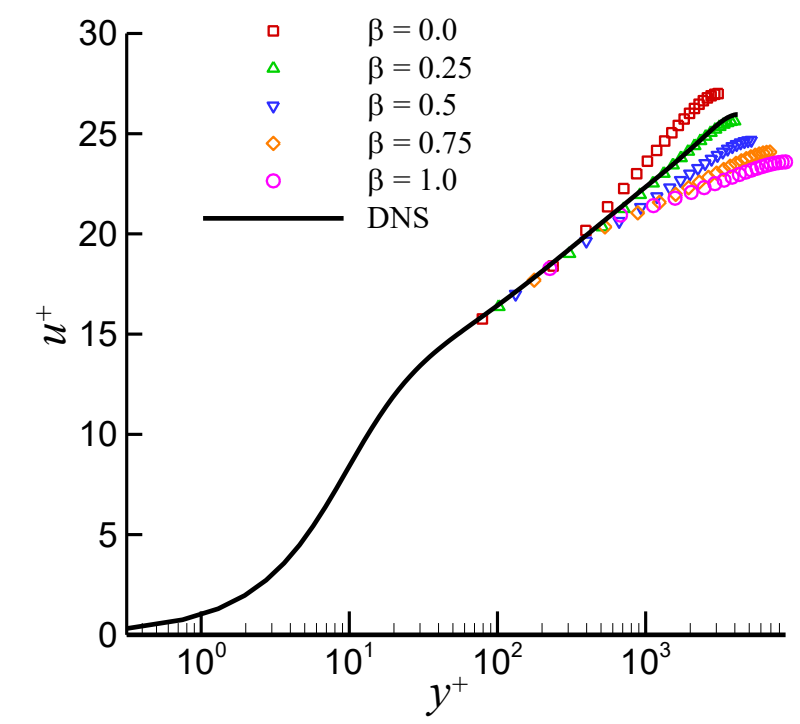

(c)

Figure 5. Influence of the blending parameter $\beta$ on the results for turbulent channel flow at target Reynolds $Re_\tau = 4200$. (a) Convergence of output friction velocity. The target friction velocity is shown in black. (b) Mean velocity profiles normalized with the target $u_\tau^*$ (c): semi-log plot of the mean velocity profiles normalized with the computed $u_\tau$

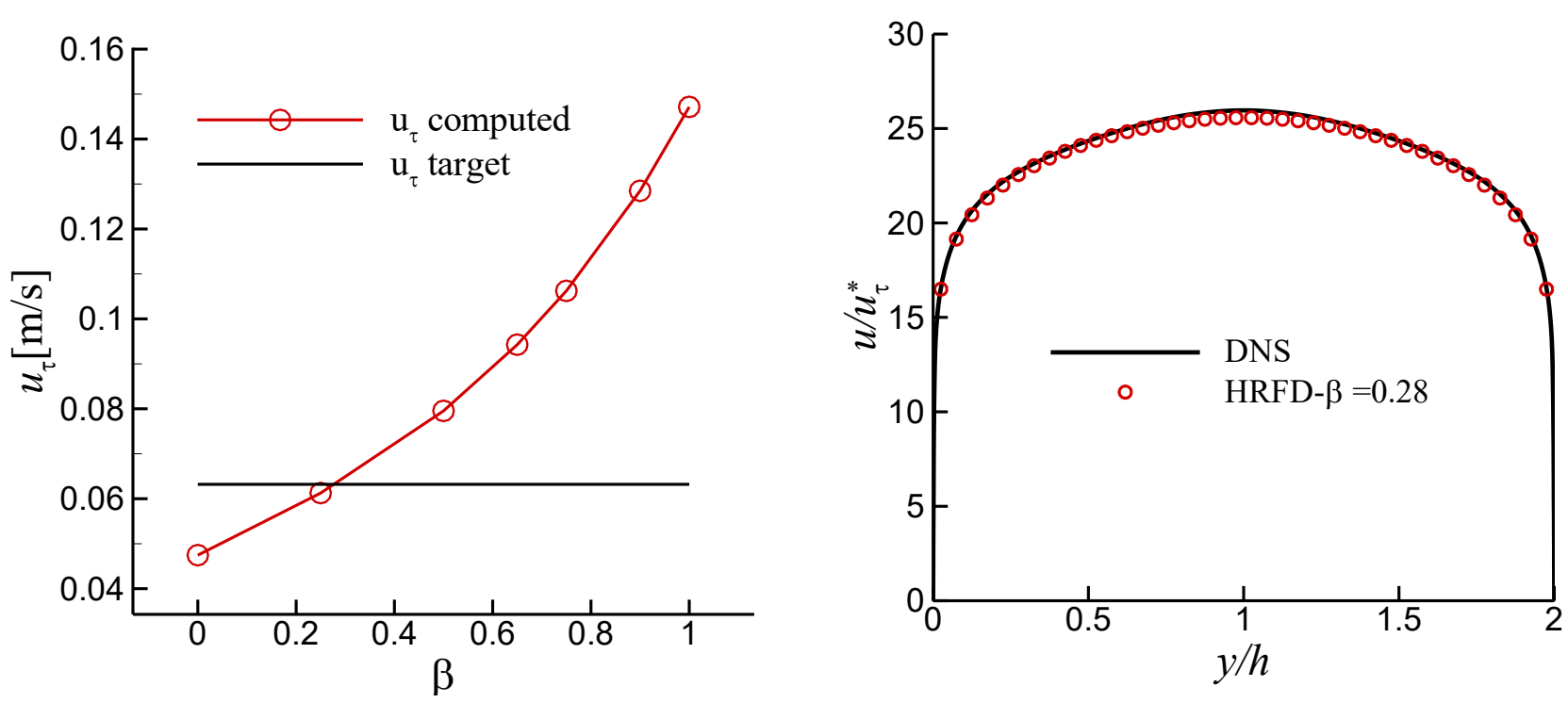

(a) (b)

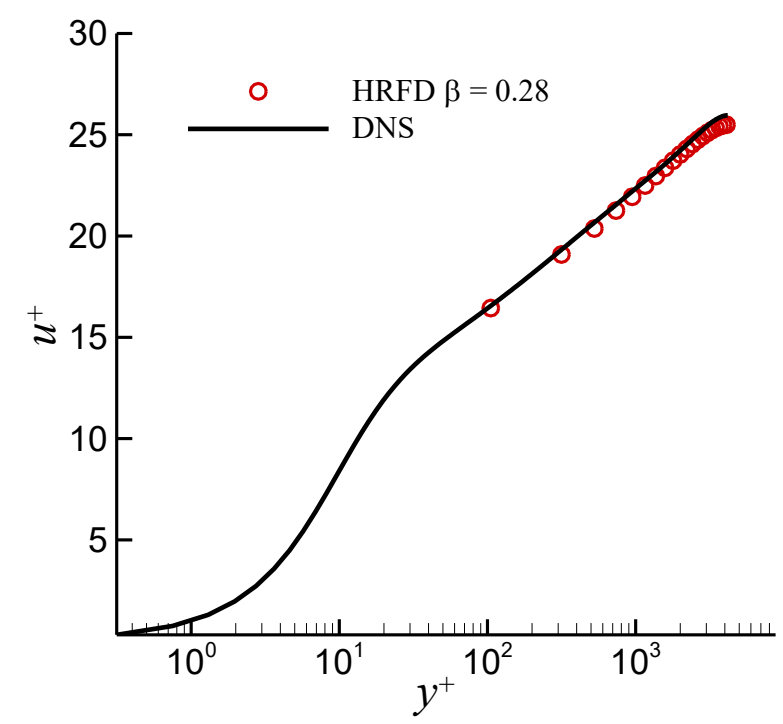

(c)

Figure 6. Calibration of $\beta$ against the target $u_\tau$ (solid black line) for turbulent channel flow at $\mathrm{Re}_\tau$ = 4200. (a) Variation of the computed $u_\tau$ for different values of the blending parameter. (b) Mean velocity profile (normalized by target $u_\tau^*$) obtained with $\beta = 0.28$ compared to DNS *[43]*. (c) semi-log plot of the velocity profile normalized by the computed $u_\tau$ compared to DNS results.

**4.1.2.2 Analysis of the Cai-Degrigny gradient reconstruction**

We now analyze the Cai–Degrigny gradient reconstruction method [21] introduced in Section 3.1.2. Figure 7 shows a comparison between the mean velocity profiles obtained using the HRFD methodology and those resulting from the Cai–Degrigny reconstruction. The results demonstrate that the Cai–Degrigny approach yields a marked improvement over the purely analytical gradient reconstruction based on the wall function ( $\beta$ = 1.0) which is not included in Figure 7 but can be seen in Figure 5(b). Moreover, the velocity profile obtained using the Cai–Degrigny method exhibits a reasonable agreement with the DNS reference data, notably without the need for any explicit calibration procedure, as is required in the HRFD approach.

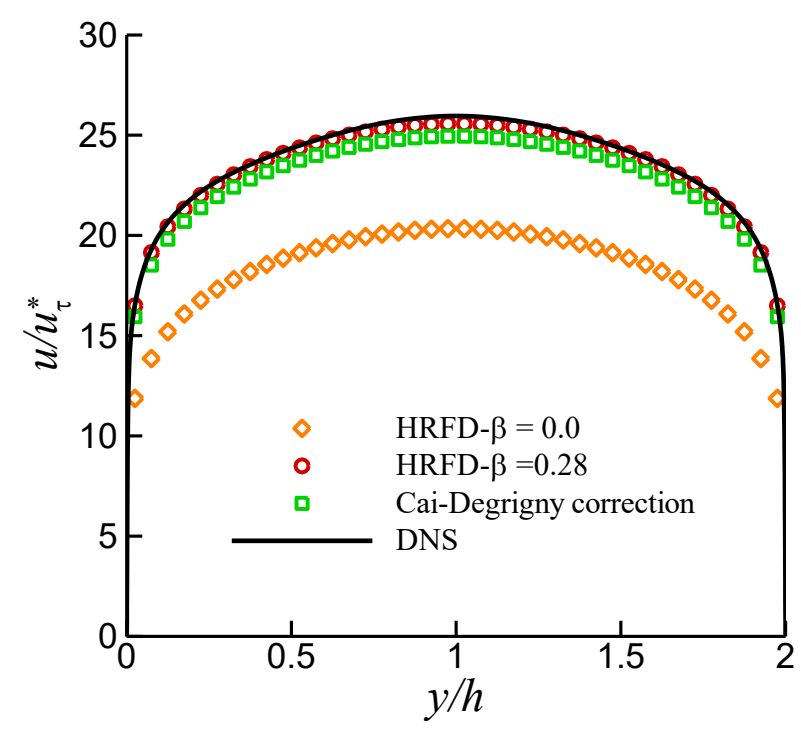

Figure 7. Mean velocity profile for regularized boundary conditions with different wall normal velocity gradient reconstruction prescriptions.

To gain further insight into the behavior of the Cai–Degrigny gradient correction and to compare it with the HRFD methodology, we evaluate the ratio between the numerical derivative obtained from Equation (31) and the analytical wall-function-based derivative from Equation (17), both computed at the boundary node. Since the Cai–Degrigny correction depends solely on the analytical velocity profile and grid spacing (i.e., it is independent of any flow solution output), this ratio can be computed a priori, without relying on LBM channel flow simulations. Figure 8 shows the results of this ratio over a broad range of near-wall grid spacings. The curve was constructed by varying the grid resolution from $N = 5$ to $N = 40$ points per half channel, for two representative friction velocity values corresponding to $Re_\tau = 4000$ and $Re_\tau = 20000$, thereby covering a wide span of $y_B^+$ and grid resolution values. The results reveal that the Cai–Degrigny correction systematically reduces the magnitude of the analytical derivative by a nearly constant factor in the range [0.76, 0.77], such that $\delta_y^{(3)} u_{WF}\big|_B \approx 0.76\,\partial_y u_{WF}\big|_B$. This behavior suggests that the Cai–Degrigny method implicitly compensates for the overestimation introduced by the raw analytical derivative, which (as shown in Figure 5) leads to inaccurate results when used directly in wall-normal gradient reconstructions. This insight, together with the fact that Cai et al. [21,32] do not provide a detailed derivation of their gradient reconstruction technique, leads us to conjecture that their prescription has been possibly inferred through numerical experimentation.

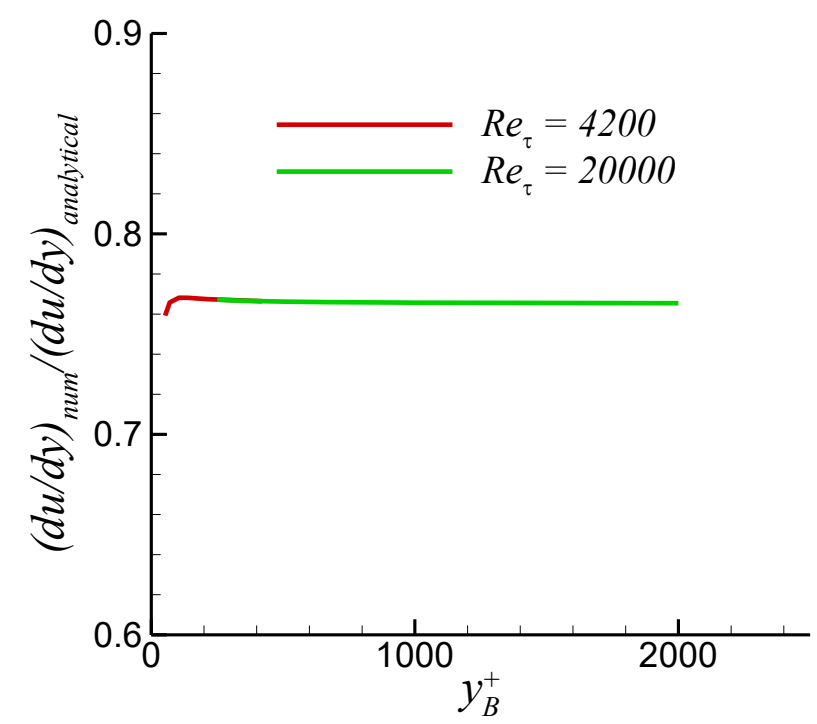

Figure 8. Ratio of the numerical derivate computed via the Cai-Degrigny approach (31) to the analytical derivate of the wall function evaluated at the boundary node.

To conclude the comparison between the HRFD and Cai–Degrigny's approaches, we show that it is possible to identify a specific value of the HRFD blending parameter that closely replicates the behavior of the Cai–Degrigny correction. As shown in Figure 9, the HRFD approach with $\beta = 0.25$ yields nearly identical results as the Cai–Degrigny method in terms of both the friction velocity convergence Figure 9(a) and the mean velocity profile Figure 9(b). Although the outcomes are equivalent, the underlying philosophies of the two approaches differ significantly. The Cai–Degrigny reconstruction does not require any prior calibration and is entirely independent of the flow solution, relying only on the analytical wall function and the grid resolution. However, its accuracy is intrinsically tied to the assumptions of the wall function. In contrast, the HRFD methodology incorporates flow-dependent information and requires an a priori calibration of the blending parameter to achieve optimal results. While this introduces additional computational effort, it also offers increased flexibility and potential for improved accuracy in more complex or non-canonical flow scenarios.

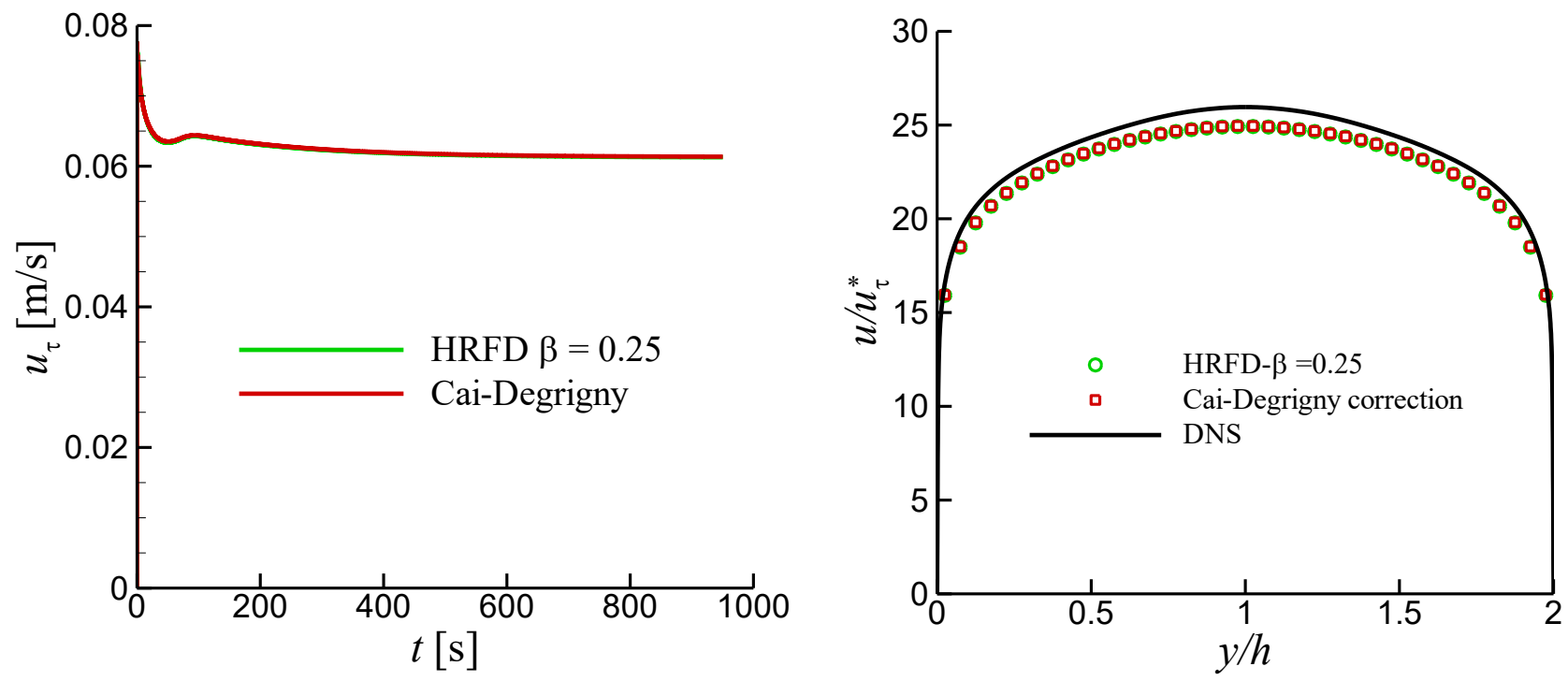

(a) (b)

Figure 9. Comparison with the HRFD approach calibrated to match the Cai-Degrigny reconstruction method (a) Convergence of $u_\tau$. (b) Mean velocity profiles normalized by the target $u^*_\tau$.

An important conclusion drawn from the numerical experiments involving regularization-based boundary schemes is that achieving reliable accuracy critically depends on the proper treatment of the wall-normal velocity component. In particular, the accuracy of the computed wall shear stress and near-wall velocity profiles is highly sensitive to the method used for reconstructing the wall-normal gradient, underscoring the need for carefully designed boundary treatments in this class of methods.

**4.1.2.3 Comparison between regularized and slip velocity bounce-back schemes**

Next, we incorporate the results obtained using the slip-velocity bounce-back scheme. Figure *10* provides a comparative summary of the previously discussed regularization-based boundary treatments (HRFD and Cai–Degrigny) alongside the slip-velocity bounce-back approach described in Section 3.2. The results indicate that the slip-velocity bounce-back scheme yields highly accurate predictions in both computed friction velocity (Figure *10*(a)) and mean velocity profile (Figure *10*(b)), without requiring any prior calibration or ad hoc treatment of the wall-normal velocity gradient. In contrast, the Cai–Degrigny method slightly under-predicts the velocity profile compared to the bounce-back result, primarily due to a small underestimation of the friction velocity. The HRFD methodology, benefiting from its prior calibration against DNS results, produces results that are comparable in accuracy to the bounce-back approach. Finally, the uncorrected regularized scheme ($\beta$ = 0) exhibits the lowest accuracy among the methods considered, which can be attributed to its inadequate representation of the wall-normal velocity gradient at the boundary node.

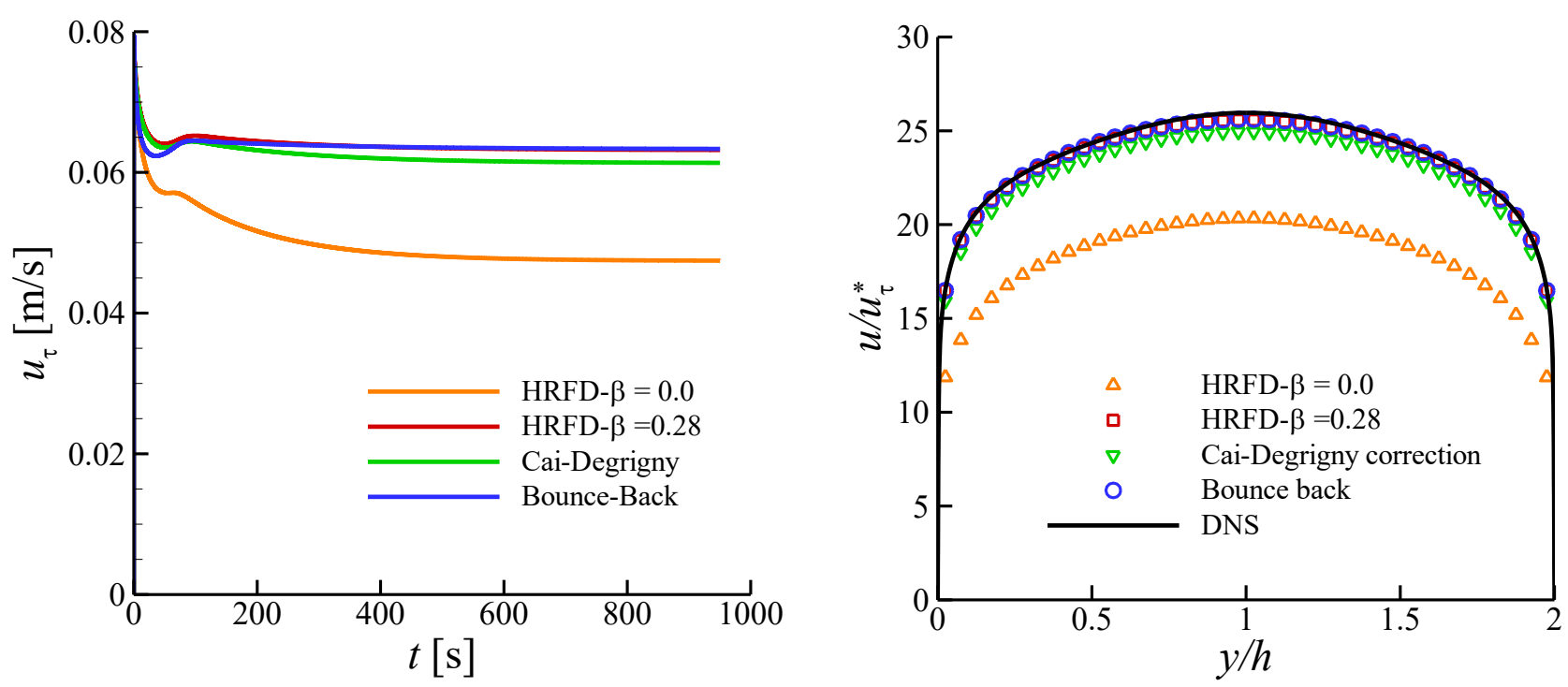

(a) (b)

Figure 10. Comparison between the bounce-back scheme and the regularized-based schemes. (a) Convergence of the friction velocity. (b) Mean velocity profiles normalized by the target $u_\tau^*$.

The numerical experiments presented above demonstrate that the accuracy of the predicted friction velocity and the corresponding mean velocity profile is directly linked to each boundary scheme's capability to accurately reconstruct the wall-normal velocity gradient at the boundary node. Figure *11* presents a comparison of wall-normal velocity gradients, post-processed from the previously shown numerical solutions using a second-order finite difference scheme. The DNS data is included as a reference. It is immediately evident that the boundary node lies within a region characterized by a steep velocity gradient, making its accurate reconstruction critical. Among the examined methods, the regularization-based schemes (specifically the HRFD and Cai–Degrigny approaches) yield reasonable approximations to the DNS gradient. In contrast, the pure regularized approach ($\beta$ = 0), previously used in laminar flows [30], and the direct use of the analytical wall-function derivative ($\beta$ = 1) result in significant overestimations or underestimations of the wall-normal velocity gradient. This inaccuracy translates into the poorest performance in terms of predicted friction velocity and mean velocity profiles, as observed in earlier figures.

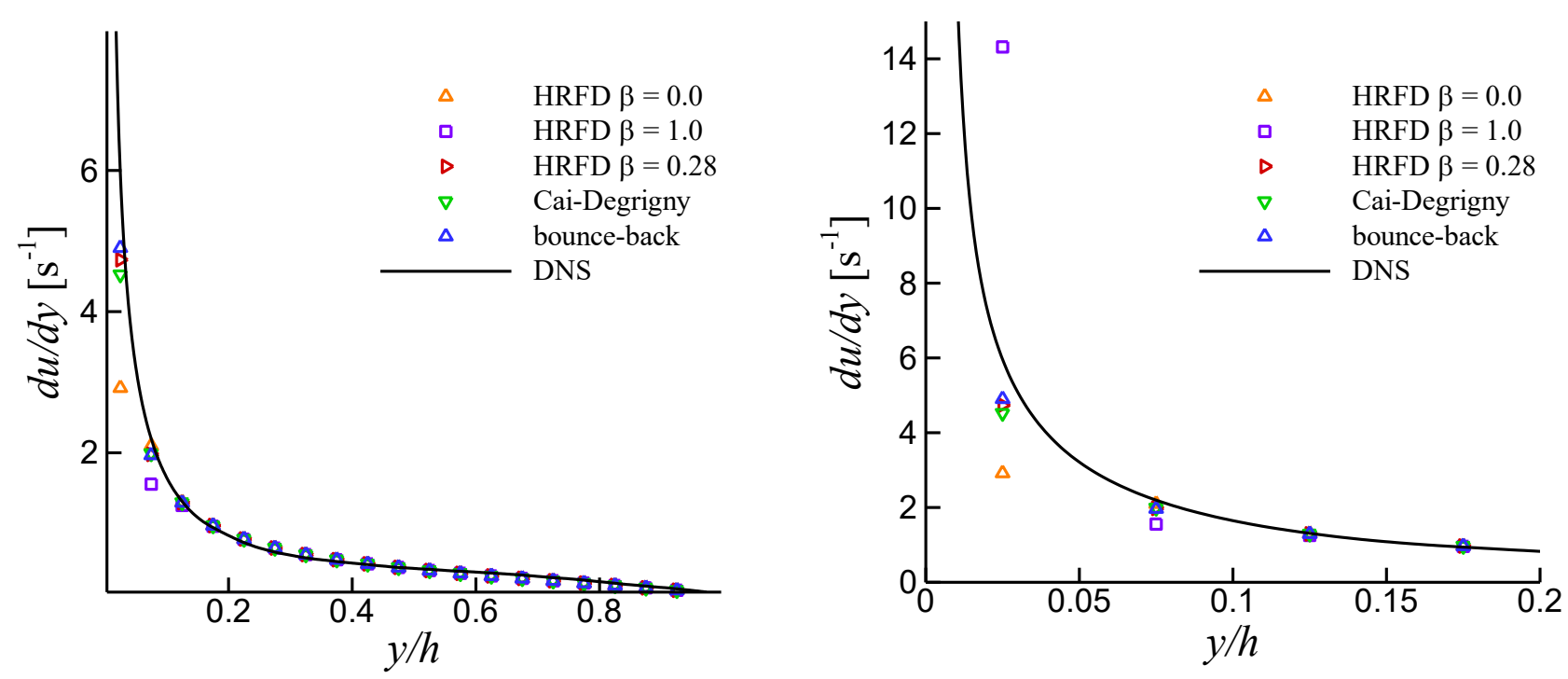

(a) (b)

Figure 11 (a) Wall normal velocity derivate obtained from the flow solution for the investigated boundary schemes. (b) Zoom of the near wall region

On the other hand, the bounce-back scheme provides the closest agreement with the DNS wall-normal velocity gradient, confirming its effectiveness in capturing near-wall turbulent behavior without requiring additional corrective procedures.

As explained above, providing a theoretical explanation for the observed numerical behavior of the different boundary schemes is a highly nontrivial task. The fundamental differences in the underlying principles of the regularized and bounce-back approaches would necessitate a rigorous multi-scale Chapman–Enskog expansion, along with a detailed analysis of the coupling between the wall model and the turbulence closure (e.g., RANS) employed in the simulation. Such an in-depth theoretical investigation lies beyond the scope of the present work and is left for future studies.

#### 4.1.2.4 Effect of grid resolution on the wall boundary schemes

Assessing the influence of grid resolution on the predicted friction velocity and mean velocity profile is critically important when employing turbulence models in conjunction with wall-function-based boundary treatments. In the context of LBM simulations at high Reynolds numbers, near-wall modeling, such as the RANS with wall functions approach employed in this work, is essential, making the evaluation of boundary scheme performance under varying grid resolutions a key aspect of model evaluation.

This issue is particularly relevant in large-scale 3D turbulent flow applications, where the computational LBM grid can easily exceed $10^8$ nodes [12]. In such cases, reducing the number of nodes within the boundary layer, without substantially compromising the accuracy of the solution, is crucial for saving computational resources. Consequently, evaluating the performance of various boundary condition schemes on under-resolved meshes is of great practical importance.

To this end, a grid convergence study has been performed for a turbulent channel flow at a reference friction Reynolds number $Re_\tau = 4200$ considering the three boundary treatments analyzed previously. Table 3 summarizes the grid resolutions studied, characterized by the number of nodes per half channel, along with the representative $y^+$ values at the boundary node and at a reference location, based on the target friction velocity.

Table 3. Grid resolutions and computed $y^+$ with the target $u_\tau$ for $Re_\tau = 4200$

| N | $y_B / h$ | $y_B^+$ | $y_F^+$ |
|---|---|---|---|
| **5** | 1.00×10$^{-1}$ | 420 | 1260 |
| **10** | 5.00×10$^{-2}$ | 210 | 630 |
| **20** | 2.50×10$^{-2}$ | 105 | 315 |
| **30** | 1.67×10$^{-2}$ | 70 | 210 |
| **40** | 1.25×10$^{-2}$ | 52.5 | 157.5 |

Figure *12* illustrates the convergence behavior of the predicted $u_\tau$ values and corresponding mean velocity profiles for the different boundary schemes under varying grid resolutions. The results show that the slip-velocity bounce-back scheme (middle row of Figure *12*) is remarkably robust with respect to grid coarsening, maintaining good accuracy even at very low resolutions (e.g., $N = 5$ points per half channel). This insensitivity to grid refinement may be attributed to the small magnitude of the coefficient $E(\tau)$ associated with the $O(\Delta^2)$ term in the multi-scale expansion (37), which limits its impact on the accuracy of the boundary condition. In contrast, the hybrid regularized finite-difference (HRFD) method (upper row of Figure *12*) exhibits a clear dependence on grid resolution. While accurate results are achieved for moderate to fine grids, the performance deteriorates significantly on coarse grids, suggesting that the calibration of the blending parameter depends on the grid resolution and is therefore not universal. This degradation is likely due to the second-order discretization error introduced when estimating the wall-normal gradient from the flow field. The Cai–Degrigny reconstruction (lower row of Figure *12*), although based on a third-order finite difference approximation, also displays sensitivity to coarse grid resolutions.

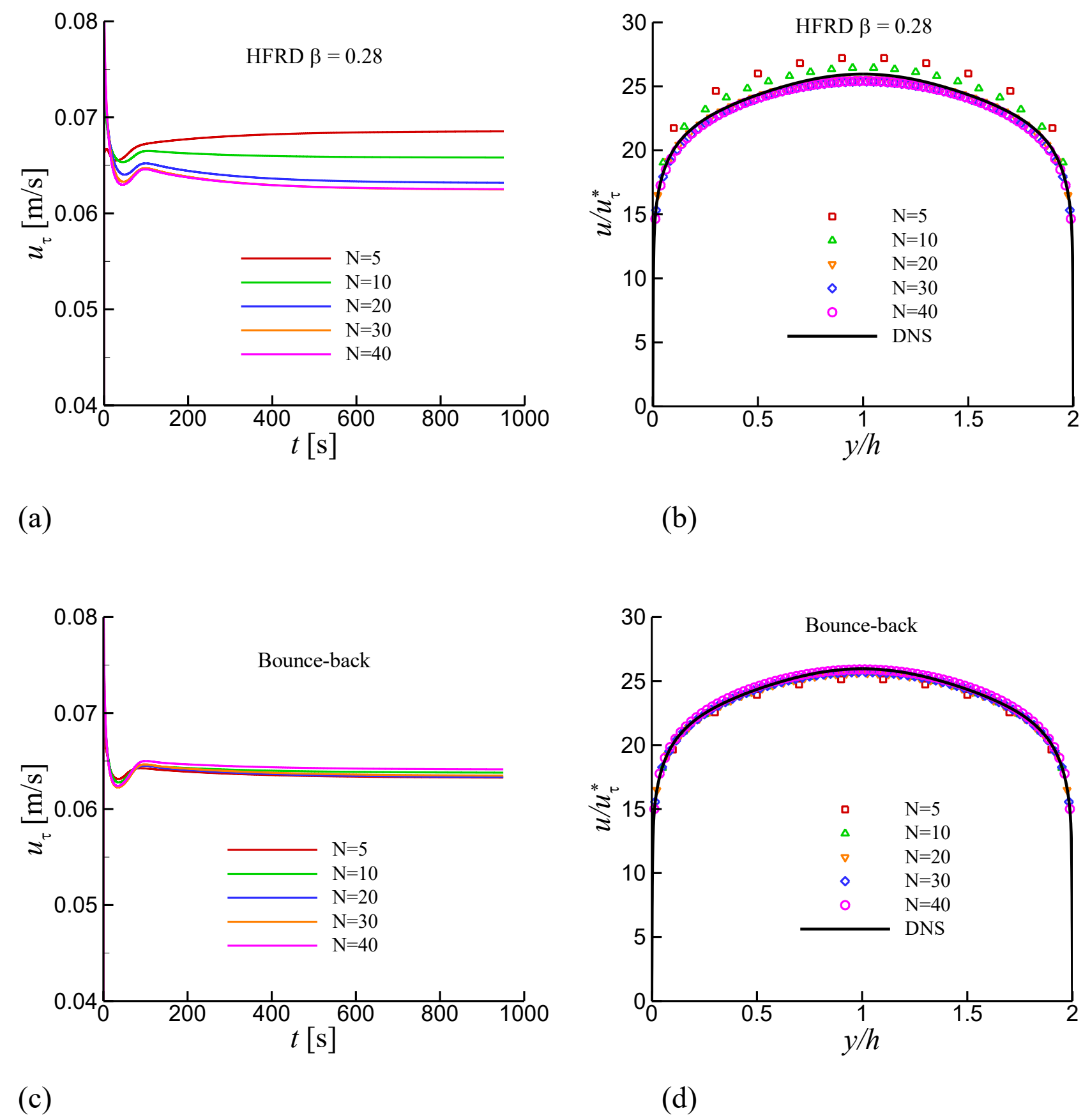
HFRD β = 0.28
$u_\tau$ [m/s]
$t$ [s]
N=5
N=10
N=20
N=30
N=40
(a)
HFRD β = 0.28
$u/u_\tau^*$
$y/h$
N=5
N=10
N=20
N=30
N=40
DNS
(b)
Bounce-back
$u_\tau$ [m/s]
$t$ [s]
N=5
N=10
N=20
N=30
N=40
(c)
Bounce-back
$u/u_\tau^*$
$y/h$
N=5
N=10
N=20
N=30
N=40
DNS
(d)

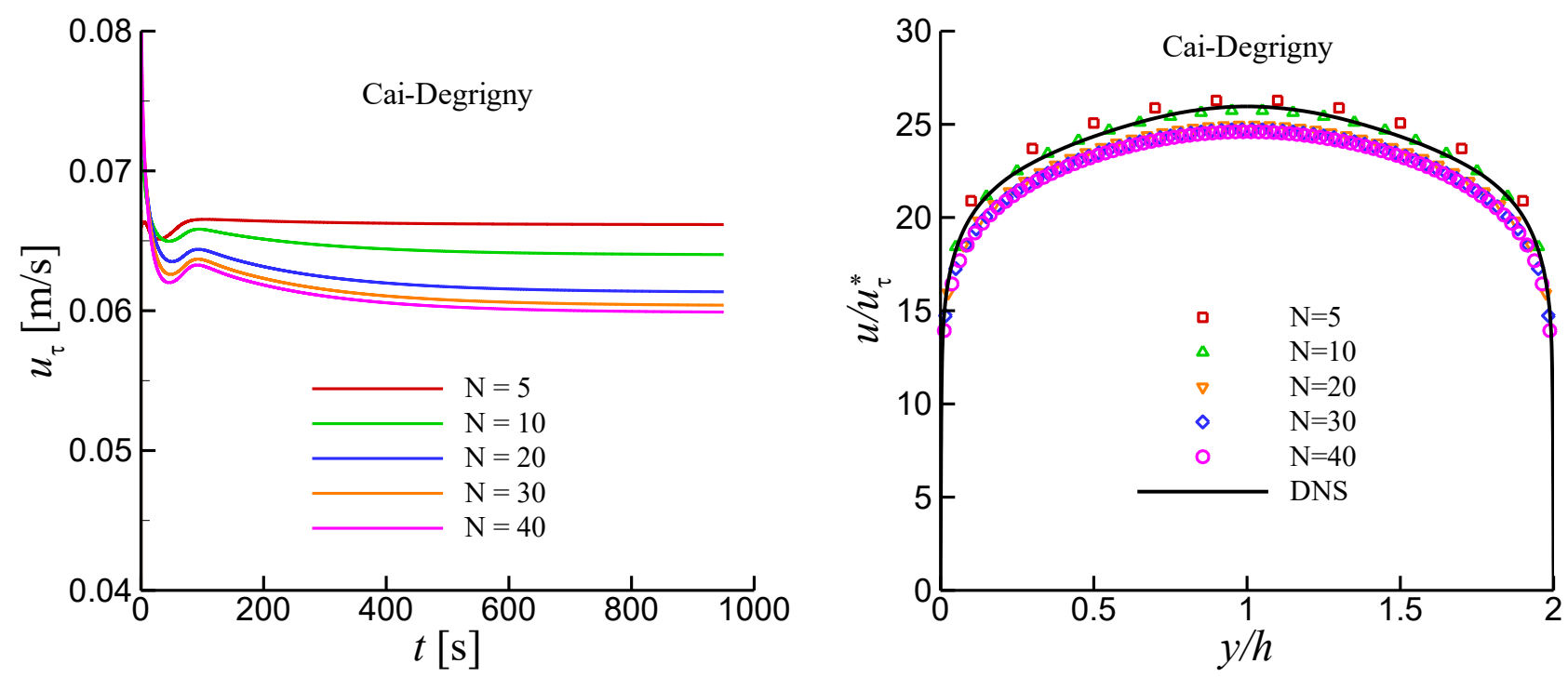

(e) (f)

Figure 12 Mesh resolution study of the different boundary schemes. (a-c-e): convergence of friction velocities. (b-d-f): mean normalized velocity profiles. Panels a & b: HRFD $\beta = 0.28$. Panels c & d: Bounce-back. Panels e & f: Cai-Degrigny scheme.

Figure 13 summarizes the convergence behavior of the predicted friction velocity $u_\tau$ as a function of the grid spacing $\Delta$. The results show a nearly linear convergence trend for the regularization-based boundary schemes (HRFD and Cai–Degrigny). In contrast, the slip-velocity bounce-back approach exhibits negligible sensitivity to $\Delta$, as previously discussed. This behavior reinforces the robustness of the bounce-back scheme under grid coarsening and further substantiates the hypothesis that its weak sensitivity to grid resolution stems from the small amplitude of higher-order error terms in the Chapman–Enskog asymptotic expansion, as outlined in Section 3.2.

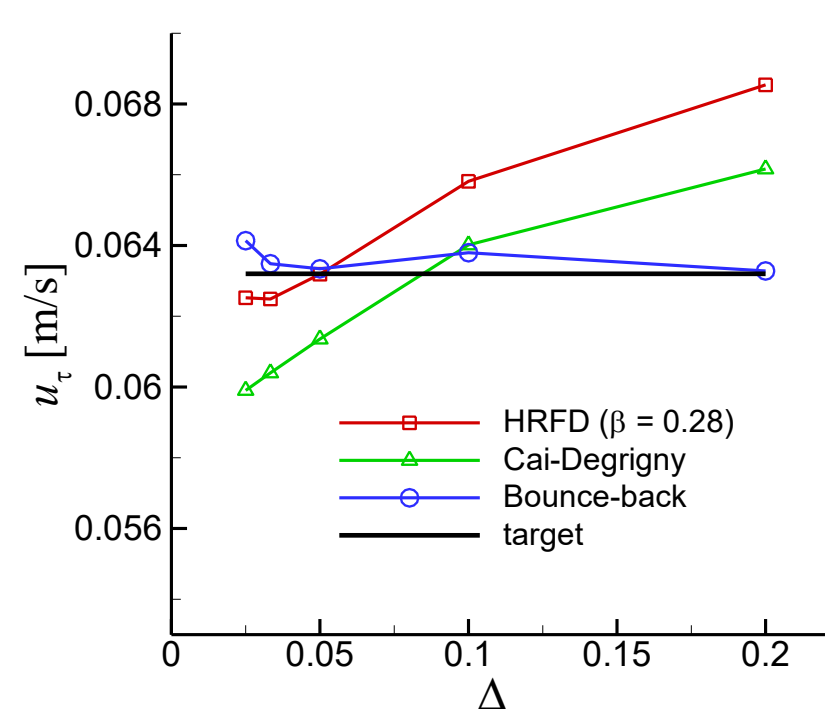

Figure 13 Grid convergence of the friction velocity for the boundary schemes

#### 4.1.2.5 Sensitivity to boundary node placement

In the results presented thus far, the boundary node was located precisely at the midpoint between two lattice nodes, i.e., at a distance $\Delta/2$ from the wall, where $\Delta$ denotes the lattice spacing. In this section, we wish to investigate the behavior of the different schemes when the wall does not lie midway between grid points. Under this configuration, interpolated bounce-back schemes, such as those proposed by Bouzidi and Mei et al., are required, as discussed in Section 3.2, so this analysis is particularly relevant for validating these schemes.

To explore the effect of the boundary node placement, we adopt Bouzidi et al. interpolated bounce-back scheme [15] and introduce a dimensionless parameter $d_w = \Delta_w / \Delta$ (as defined in Figure 2(b)) to characterize the normalized distance from the wall to the first fluid node. A parametric study is conducted by varying $d_w$ from 0.3 to 0.7 for the three boundary schemes previously analyzed, using a turbulent channel flow at $Re_\tau = 4200$ and a fixed grid resolution of $N = 20$ points per half channel. Figure 14 summarizes the results obtained from these simulations. A lower bound of $d_w = 0.3$ was fixed to avoid placing the boundary node within the buffer layer, a region where the assumptions of the wall function approach may no longer be valid. Accordingly, this study focuses on evaluating the performance of the boundary schemes in the log-law-dominated region, where wall modeling is most relevant and meaningful.

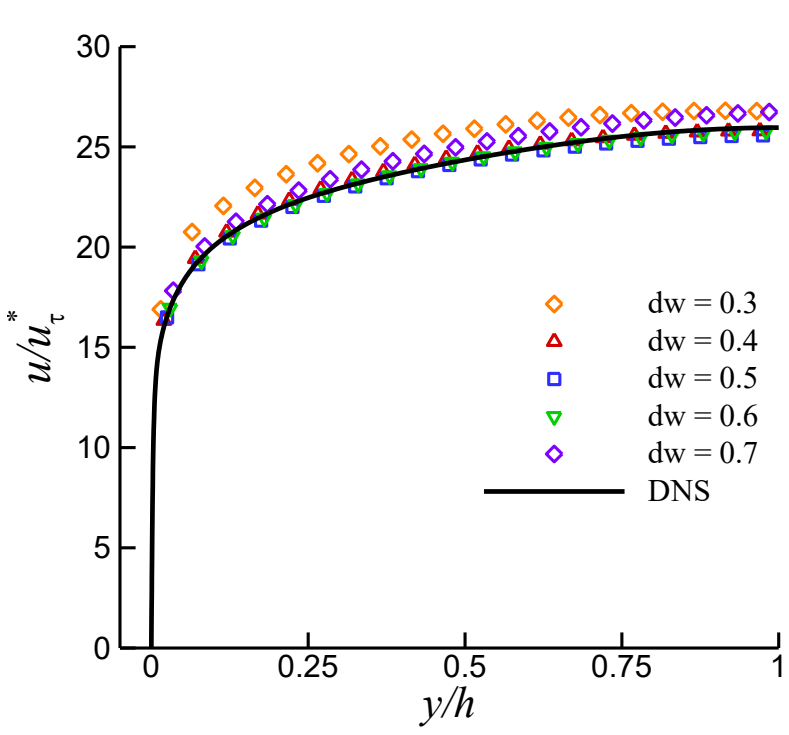

(a)

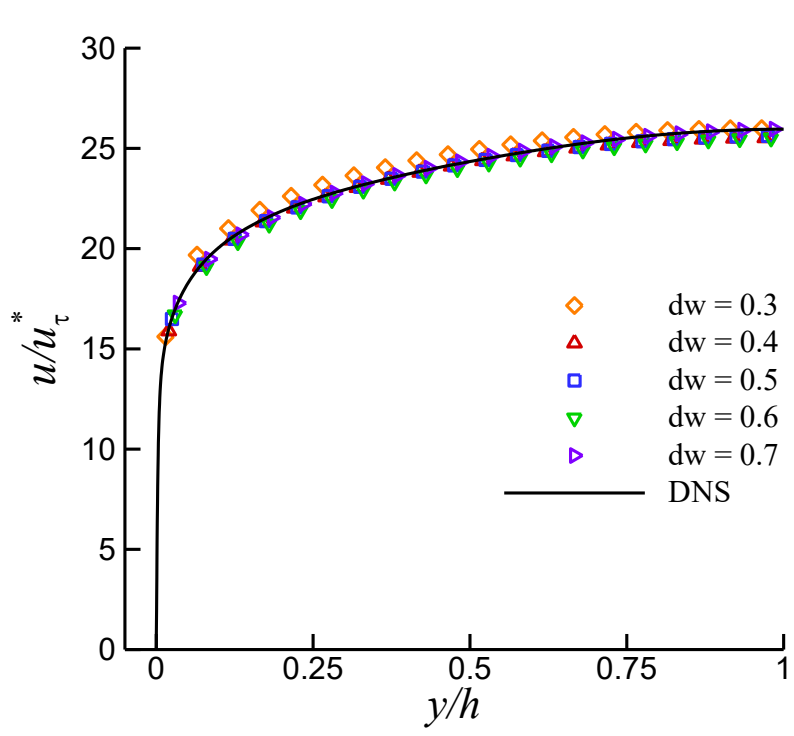

(b)

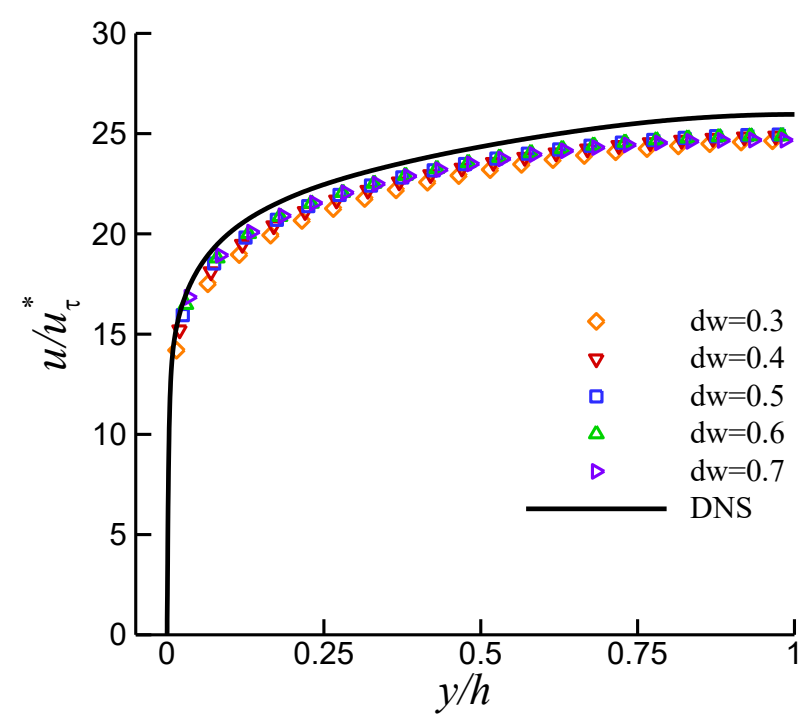

(c)

Figure 14. Mean velocity profiles for different wall distances of the boundary node represented by the parameter $d_w$. (a): HRFD $\beta = 0.28$ (b): slip velocity bounce-back (c): Cai-Degrigny gradient reconstruction

The results displayed in Figure 14 reveal that the HRFD approach is the most sensitive to the wall-normal location of the boundary node, particularly when the node is positioned close to the wall (i.e., for lower values of $d_w$). This behavior is expected, as the velocity gradient becomes increasingly steep near the wall, amplifying the discretization error associated with the numerical estimation of the wall-normal derivative from the flow field. As a result, the HRFD scheme exhibits notable degradation in accuracy under these conditions.

In contrast, the slip-velocity bounce-back scheme shows minimal sensitivity to the boundary node placement. Only for the smallest values of $d_w$ a slight deviation of the velocity profile is observed, but the differences remain within acceptable bounds. This confirms that the bounce-back approach maintains its robustness and accuracy even when implemented with interpolated schemes, such as that of Bouzidi et al.

Finally, the Cai–Degrigny gradient reconstruction approach demonstrates a high degree of insensitivity to the distance between the boundary node and the wall. For the grid resolution considered, its performance remains essentially unchanged across the range of wall-normal distances tested, further reinforcing the robustness of this method with respect to node positioning in the near-wall region.

#### 4.1.2.6 Reynolds number validation for the channel flow

To conclude the analysis of the turbulent channel flow test case, we examine the influence of the friction Reynolds number $Re_\tau$ on the predicted friction coefficient for the three boundary condition schemes discussed previously. The empirical correlations of Dean [42] and Petukhov [44] have been employed as reference experimental correlations for validation purposes. Figure 15 presents the variation of the bulk friction coefficient $Cf_b = 2(u_\tau / U_b)^2$ as a function of the bulk Reynolds number $Re_b$ , for a fixed grid resolution of N = 20. The test cases span a range of friction Reynolds numbers from $Re_\tau = 2\times10^3$ to $Re_\tau = 2\times10^4$. The results show that all three boundary schemes reproduce the expected trend of the friction coefficient with increasing Reynolds number reasonably well, and remain consistent with the reference empirical correlations across the tested range. It is worth noting that normalizing the skin friction by the bulk velocity reduces discrepancies between different boundary schemes compared to direct comparisons based on friction velocity $u_\tau$. In addition, the semi-logarithmic plots of the corresponding mean velocity profiles are shown in Figure *16*. These results indicate that, at this grid resolution, all three boundary schemes accurately capture the logarithmic layer of the mean velocity profile across the full spectrum of Reynolds numbers considered. This confirms the capability of each approach to represent turbulent wall-bounded flows over a broad range of flow regimes when used in conjunction with a RANS-based wall function model.

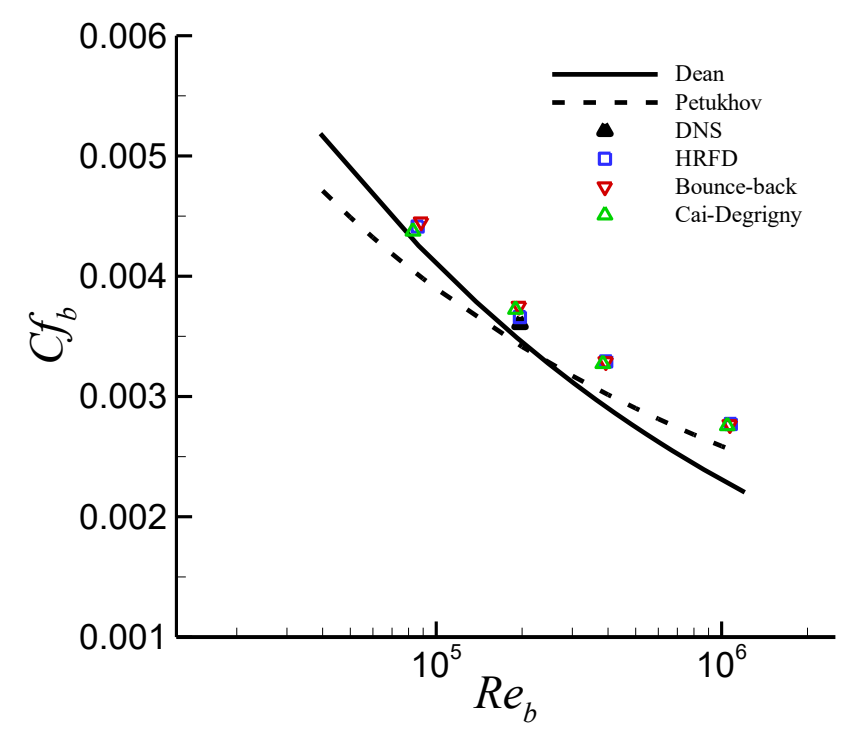

Figure 15 Bulk friction coefficient with respect to the mean Reynolds number computed for the three boundary schemes. The plain and dashed lines represent the experimental Dean and Pethukov correlations.

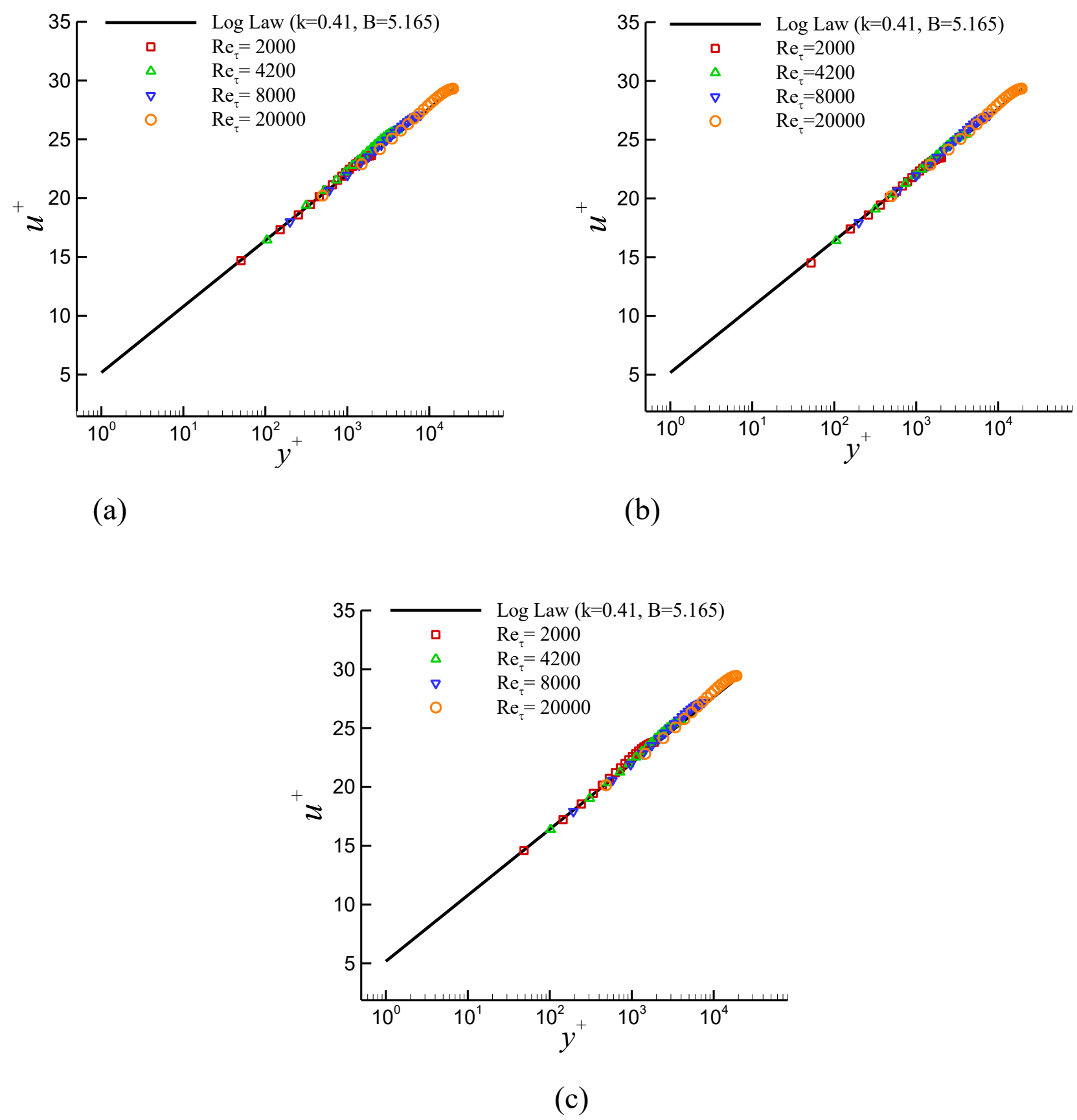

Figure 16. Log-law velocity profiles for the different boundary schemes. (a) HRFD $\beta = 0.28$, (b): bounce-back (c) Cai-Degrigny gradient reconstruction

## 4.2 Zero pressure gradient turbulent flat plate

The flat plate test case is taken from NASA's Turbulence Modeling Resource (TMR) website [39] and serves as a benchmark for external aerodynamic flows. Despite its geometric simplicity, this test case provides a meaningful verification and validation scenario to assess the performance and robustness of the implemented boundary condition strategies. If the boundary treatment schemes fail to accurately predict the flow over a flat plate (characterized by zero pressure gradient and simple geometry) they are unlikely to perform reliably in more complex configurations involving surface curvature or adverse pressure gradients, which pose significant challenges for near-wall modeling when using wall functions.

LBM simulations are carried out using the SA turbulence model in conjunction with the wall function approach described in Section 2 and are compared against reference results obtained with NASA's CFL3D solver. This code solves the RANS equations using a second-order finite volume, cell-centered discretization on structured grids [45] and uses the same version of the SA model used by our LBM code.

### 4.2.1 Test case description

This test case involves a zero-pressure-gradient turbulent boundary layer developing over a flat plate of length 2. The simulation is performed at a nominal Reynolds number $Re$ = $5\times10^6$ based on a unit reference length $L_{ref}$ = 1 m, resulting in a local Reynolds number of $Re = 10^7$ at the downstream edge of the plate. Figure *17* illustrates the computational domain and the prescribed boundary conditions used for the simulation setup.

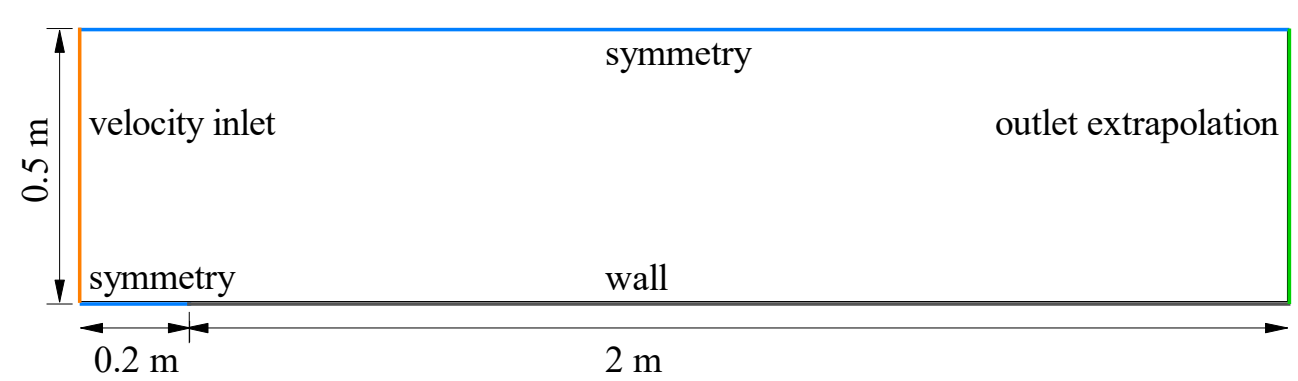

Figure 17. Basic layout of the computational domain with boundary conditions

The computational domain has length $L_x / L_{ref} = 2.2$ and height $L_y / L_{ref} = 0.5$. The flat plate starts at $x / L_{ref} = 0.0$ and the inlet boundary is located at $x / L_{ref} = -0.2$. The domain's vertical size, $L_y / L_{ref}$ was chosen to optimize computational resources without significantly compromising the accuracy. In fact, a test in which the height was doubled only changed the skin friction at $x / L_{ref} = 0.97$ by less than 0.4%. The LBM computations have been carried out on a series of progressively refined uniform Cartesian meshes (see Table 4). Flow conditions are inlet/free-stream velocity, $U = 75\,m/s$, free-stream viscosity $\nu = 1.5\times10^{-5}\,m^2 \times s$ and free-stream Mach number $M$ = 0.1. An acoustic time step is prescribed for the LBM simulation. The boundary conditions applied for this test case are summarized in Figure 17. At the inlet (left boundary), a velocity boundary condition is imposed using the Zou–He method [46], while the outlet (right boundary) employs a population extrapolation strategy following [47]. A symmetry boundary condition is enforced at the top boundary. Along the bottom boundary, the treatment is split into two regions: from the right domain boundary up to the leading edge of the flat plate (located at $x / L_{ref}$ =0.0), a symmetry boundary condition is used to mimic the free-slip surface upstream of the plate. Downstream of this point, along the flat plate surface, the turbulent wall boundary condition is applied using one of the three methods investigated: the hybrid regularized finite-difference (HRFD) scheme, the regularized approach with Cai–Degrigny gradient reconstruction and the slip velocity bounce-back method. For the HRFD method, the same blending factor $\beta = 0.28$ as determined from the channel flow test case is employed.

The comparison of boundary condition schemes has been conducted using three grid resolutions representative of the typical $y^+$ ranges encountered in wall-function-based meshes for high Reynolds number flows. Table *4* summarizes the domain sizes, mesh resolutions, and the estimated $y_B^+$ values at the boundary node near the reference station located at $x/L_{ref}$ =1. For reference, the CFL3D solution employs a highly refined mesh with $y_B^+ < 0.1$ at the wall and approximately 255 points within the boundary layer at this location. In contrast, the finest LBM mesh evaluated is over six times coarser than the CFL3D mesh at the same station, reflecting the reduced near-wall resolution characteristic of wall-function approaches.

Table 4. Grid resolutions for the flat plate test case. Note that for LBM, node *B* is located at $y = \Delta/2$ from the wall. CFL3D mesh information is obtained from the NASA Turbulence Modeling Resource *[39]*

| **Mesh** | Δ | $y_B^+$ | **Approx number of points within the TBL at *x*/*L* = 1.0** | $N_x \times N_y$ |
|---|---|---|---|---|
| **coarse** | $2\times10^{-3}$ | 180 | 10 | 1100×250 |
| **medium** | $1\times10^{-3}$ | 90 | 20 | 2200×500 |
| **fine** | $5\times10^{-4}$ | 45 | 40 | 4400×1000 |
| **CFL3D** | $5\times10^{-7}$ | 0.09 | 255 | 545×385 |

The accuracy of the LBM-RANS method using wall functions cannot be directly compared to the CFL3D RANS solution, which was obtained on a very fine mesh with a wall-resolved turbulence model, without taking into account several considerations. While the latter resolves turbulence equations up to the wall with exact boundary conditions (zero velocity and turbulent eddy viscosity), the wall-function approach relies on approximate boundary prescriptions. Moreover, near the leading edge of the flat plate, the boundary layer is too thin for wall functions to apply effectively, e.g. reference probe points may even lie outside the boundary layer. Furthermore, the grid resolution is insufficient to capture turbulence accurately compared to the fine mesh used in CFL3D. As an illustration, the number of grid points within the boundary layer at $x/L_{ref} = 0.1$ is 1, 2, and 4 for the coarse, medium, and fine LBM grids, respectively, whereas the CFL3D mesh includes 188 points at the same location. As the boundary layer grows, the wall-function approach becomes more accurate and LBM results begin to agree better with those obtained with CFL3D. Still, some differences remain due to the parabolic nature of the boundary layer equations, which makes downstream flow dependent on upstream conditions.

Taking into account these considerations, we found that an adaptation length was necessary for the wall-function approach to achieve sufficient accuracy in resolving the outer-layer region of the turbulent boundary layer and delivering reliable results. This adaptation length has been accounted for in the analysis that follows in order to enable a meaningful comparison of turbulent flow results across different boundary schemes with the CFL3D data. Specifically, since the location $x / L_{ref} = 0.97$ is the reference station selected by NASA for verification purposes [39], we aligned the LBM solution with the CFL3D data at this point. For each LBM result, we picked the location where the boundary layer and momentum thickness best match the CFL3D values at $x / L_{ref} = 0.97$. We then shifted the wall-function results so that the boundary layer thickness aligns at this reference station. (A similar procedure is carried out in [21], where velocity profiles with the same value of $Re_\theta$ are compared). In practice, this shift is small for the bounce-back boundary condition. However, due to the slower evolution of turbulence and the turbulent boundary layer under regularized boundary schemes, the extracted velocity profiles correspond to slightly downstream stations. The skin friction coefficient distributions along the flat plate have been adjusted accordingly using the same alignment criterion.

#### 4.2.2 Results

Figure 18 illustrates the mesh convergence of the skin friction coefficient distribution along the flat plate for the three boundary condition schemes and three mesh resolutions. As expected, significant discrepancies are observed near the leading edge of the flat plate across all methods, particularly for the coarse grids. This is not surprising, as the boundary layer is just beginning to develop in this region and is too thin to be accurately resolved by the wall-function approach. As the boundary layer grows downstream, the number of grid points within it increases, leading to improved resolution. Therefore, our analysis focuses on the region downstream of $x / L_{ref} = 0.95$, where the turbulent boundary layer is assumed to be sufficiently developed for the wall-function approach to yield valid results. Within this region, the three meshes span most of the $y^+$ range typically encountered in aeronautical applications. Figure 18(b) demonstrates that the slip-velocity bounce-back approach exhibits a very good grid convergence towards the CFL3D reference solution. This is particularly noteworthy given the substantial differences between the two CFD methods in terms of both physical modeling equations (kinetic vs Navier-Stokes) and numerical formulation. These results confirm that the LBM-RANS approach, when coupled with the wall-function method under equilibrium conditions, can yield highly accurate predictions of the flat plate skin friction coefficient. Furthermore, the bounce-back approach shows significantly lower sensitivity to mesh resolution compared to the regularized schemes, reinforcing observations made previously in the channel flow test case. In contrast, the regularized schemes HRFD and the Cai-Degrigny gradient reconstruction produce larger errors in the skin friction coefficient, particularly for the coarse mesh, where deviations with respect to the CFL3D results are clearly evident.

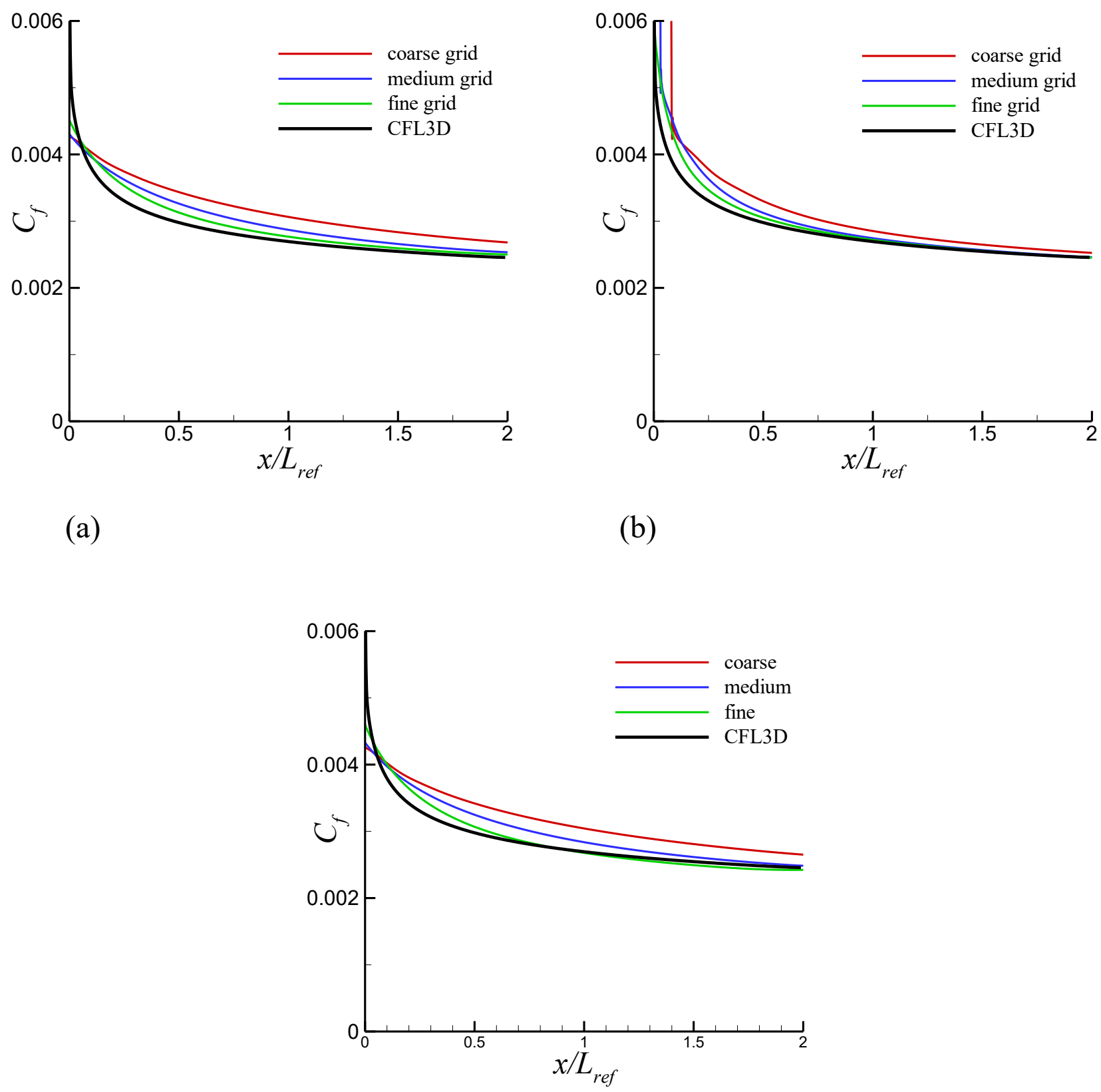

(a) (b)

(c)

Figure 18. Mesh convergence study of the skin friction coefficient ($C_f$) distribution along the flat plate for the three boundary condition schemes. Results are shown for three mesh densities (coarse, medium, fine) (a) HRFD with $\beta = 0.28$, (b) slip velocity bounce-back (c) Cai-Degrigny gradient reconstruction. For the LBM results, values of $x$ are shifted so as to match the CFL3D BL thickness at $x/L = 0.97$.

To quantify the error with respect to the reference CFL3D results without focusing on any specific station, Figure 19 compares the error in the integrated skin friction between $x = 1.0$ and $x = 2$ obtained with each boundary scheme. The error is defined as:

$$\varepsilon = \frac{\int_{x=1}^{x=2} \left| C_f^{LBM}(x) - C_f^{CFL3D}(x) \right| dx}{\int_{x=1}^{x=2} C_f^{CFL3D}(x) dx}$$

The bounce-back scheme consistently yields higher accuracy compared to the regularized-based schemes. Furthermore, it is worth emphasizing that under under-resolved conditions (common in 3D LBM simulations at high Reynolds numbers) the bounce-back approach maintains superior performance, demonstrating greater robustness and reliability in such scenarios.

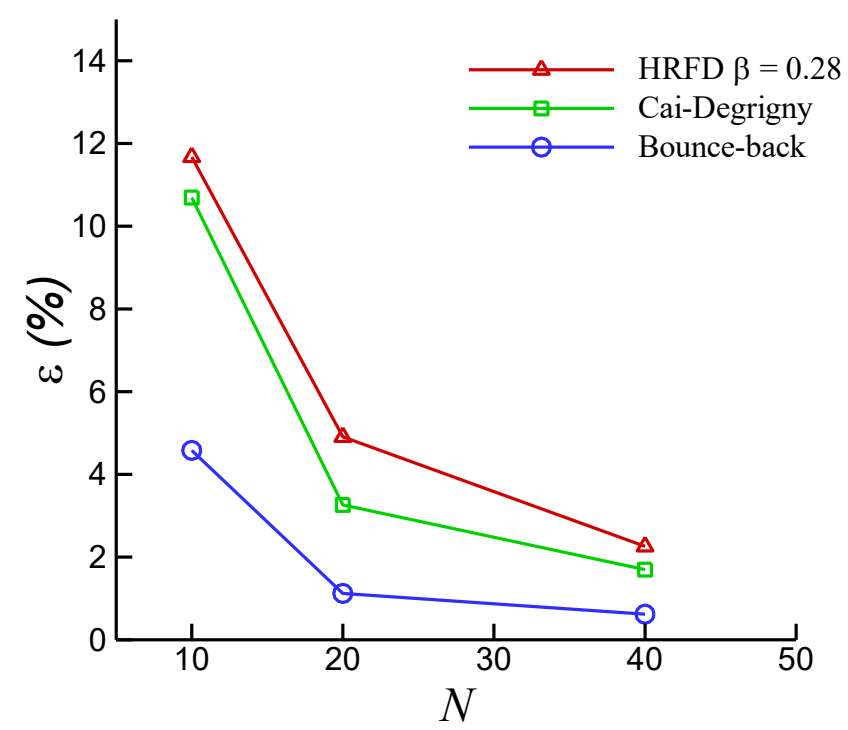

Figure 19. Relative error in % of the integrated skin friction coefficient (using trapezoidal rule) along the flat plate (from $x/L_{ref}$ = 1 to $x/L_{ref}$ = 2). The abscissae correspond to the approximate number of points within the boundary layer at $x/L_{ref}$ = 1.0 ($N$ = 10 for the coarse mesh; $N$ = 20 for the medium and $N$ = 40 for the fine mesh).

Figure 20 provides a more detailed assessment of the performance of the three boundary condition schemes analyzed in this study. The figure compares LBM results with the CFL3D reference solution at $x$ = 0.97. The top row displays the boundary layer velocity profiles, normalized by the outer flow velocity, for each boundary condition scheme. For medium and fine mesh resolutions, all three methods show very good agreement with the reference solution. However, differences become more noticeable for the coarsest mesh. In that case, the bounce-back scheme still performs remarkably well, accurately capturing the velocity profile despite the low mesh resolution.

Differences among the three boundary condition schemes become more evident in the semi log-law plots shown in the middle row of Figure 20. This is primarily due to the influence of the computed friction velocity $u_\tau$ on the scaled variables $y^+$ and $U^+$. The bounce-back scheme demonstrates better convergence and more accurate predictions compared to the regularized-based methods. The latter exhibit noticeable deviations from the CFL3D results, particularly within the defect layer region, where sensitivity to grid resolution is clearly visible.

Finally, the bottom row of Figure 20 displays the turbulent eddy viscosity profiles for each LBM boundary scheme. All three approaches provide reasonably accurate predictions, especially considering the challenges associated with simulating the flat plate case using a wall-function model, as discussed previously. Nonetheless, the Cai-Degrigny gradient reconstruction method shows more variability and deviation. In contrast, the bounce-back and HRFD methods improve their prediction of the peak eddy viscosity as the mesh is refined, a trend not observed for the Cai-Degrigny scheme.

In summary, the analysis of this benchmark test case demonstrates that the kinetic-theory-based LBM methodology represents a viable alternative to traditional Navier-Stokes solvers for simulating high-Reynolds-number turbulent flows. Despite the inherent limitations of the wall-function approach, the LBM results show very good agreement with those obtained from CFL3D simulations, which are based on a fundamentally different CFD framework, namely, the solution of RANS equations using finite volume methods without wall treatment.

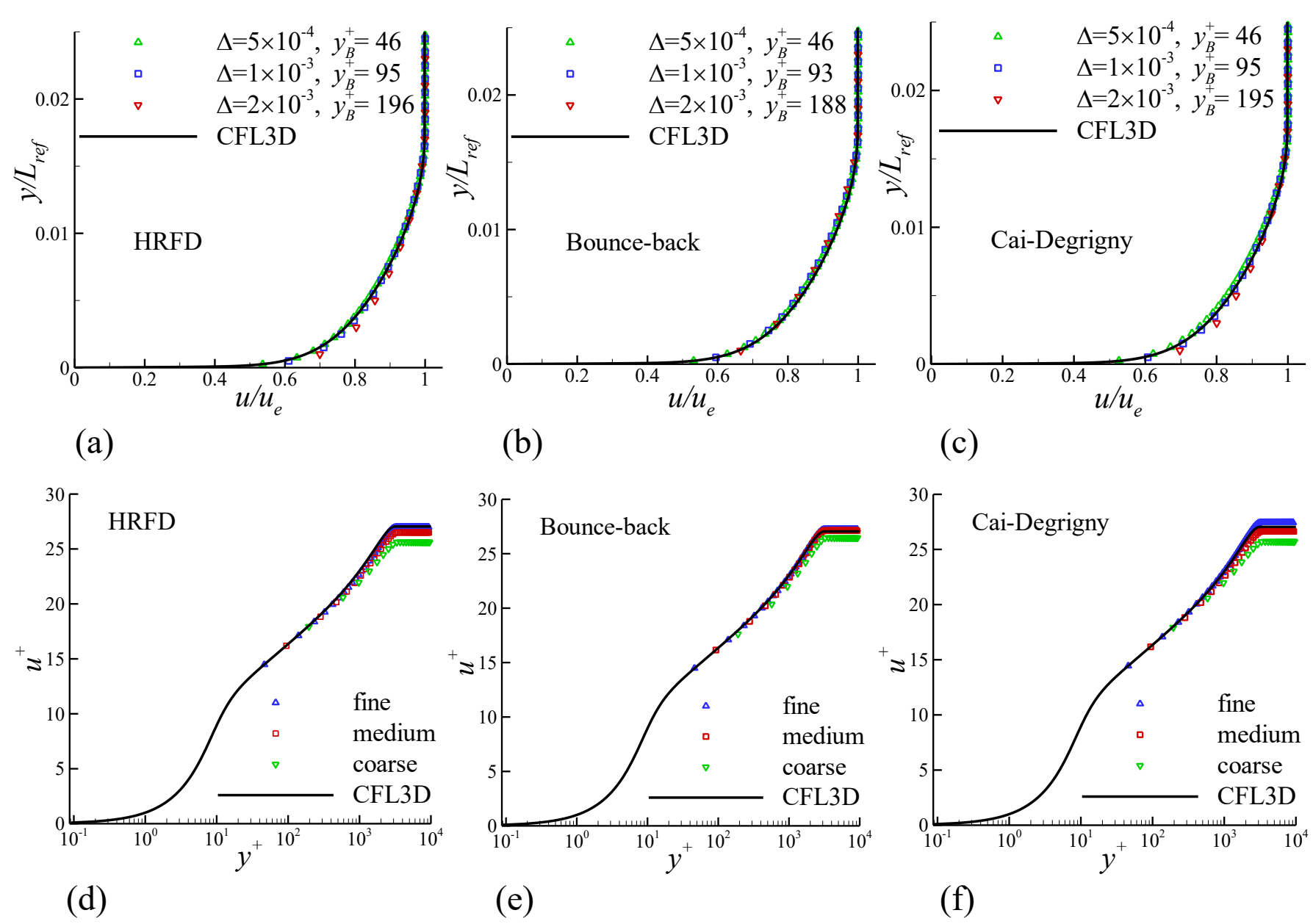

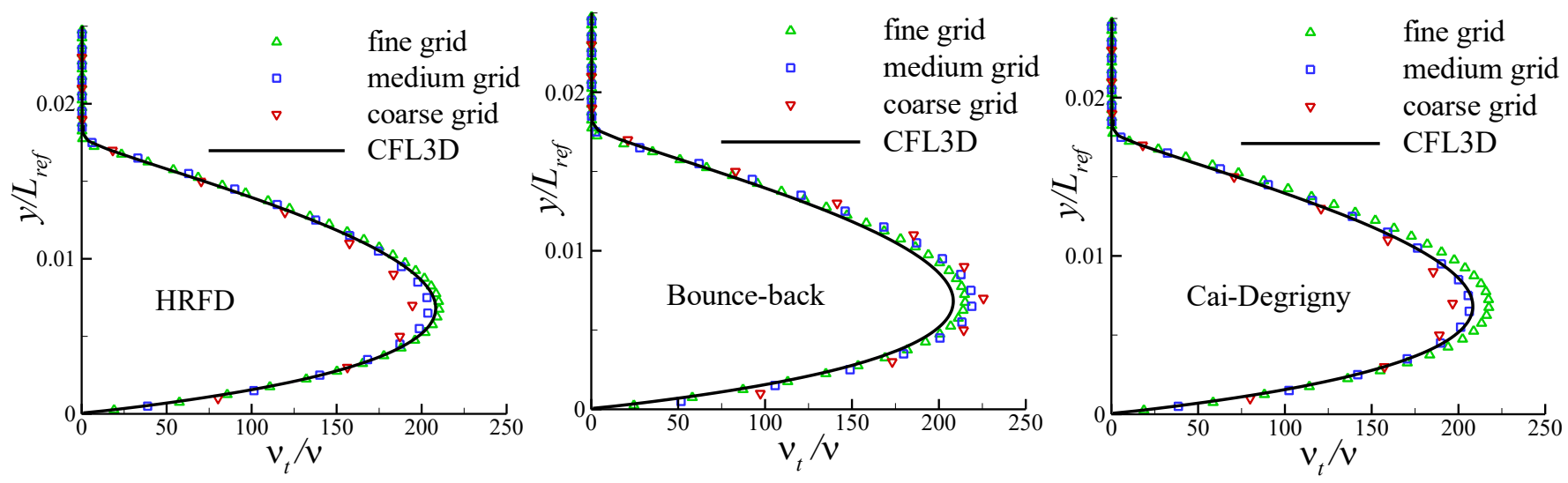

(g) (h) (i)

Figure 20. Mesh resolution study for the three boundary schemes compared to CFL3D results at $x/L_{ref}$= 0.97 on the fine (545×385) grid from NASA's Turbulence Modeling Resource [39]. The LBM results correspond to locations with the same BL thickness as CLF3D at $x/L$ = 0.97. Panels a-b-c: Streamwise velocity profiles normalized by the outer velocity. Middle row, Panels d-e-f: semi-log velocity profiles normalized by $u_\tau$. Panels g-h-i: turbulent eddy viscosity ratio profiles across the boundary layer. Panels a-d-g: HRFD with $\beta$ = 0.28. Panels b-e-h: slip velocity bounce-back. Panels c-f-i: regularized scheme with Cai-Degrigny [20] gradient reconstruction.

### 4.3 Discussion

The two types of boundary schemes described above differ fundamentally in their formulation and physical interpretation, but both yield accurate results in numerical applications, as has been shown in previous sections. In the two cases analyzed above, a pattern seems to emerge in the comparative behavior of the different boundary methods. In the simple test cases examined here, the bounce-back scheme seems to outperform the regularized schemes in overall accuracy and behavior under mesh coarsening. While the analysis is certainly limited in that only simple cases without curvature or complex flow effects have been considered and it is thus premature to extract definitive conclusions from the present work, it may be interesting to compare more closely the philosophies of each method seeking possible mechanism that explains the observed differences in behavior.

The first point concerns the impact on the results of the wall function approach or the turbulence model used in the analysis. Even though the same turbulence model and wall function are used for all three boundary schemes, a different turbulence model may interact differently with the boundary schemes and alter the conclusions. In this regard, we have repeated the analysis with the Baldwin-Lomax model and an algebraic wall function taken from [48] without significantly affecting the results. One may also wonder whether dropping the wall function assumption and fully integrating the turbulence model to the wall with sufficiently fine meshes would have some effect on the conclusions. We could expect to see an improvement on the behavior of the regularized schemes, as their chief problem is precisely their dependence on the near wall gradient computation on under-resolved meshes. However, our interest in this work is precisely in turbulent computations with wall functions, which are a must in high Reynolds number turbulent simulations in 3D or more complex configurations due to the computational cost incurred by the specific mesh requirements of the LBM.

A separate question is that of the accuracy and exactness of the boundary schemes [23], understood as the order of truncation errors and the ability to reproduce solutions of a certain polynomial order, respectively. These concepts are linked in conventional NS discretizations but diverge for LBM. The LBM is second order accurate in space in bulk and is also able to reproduce parabolic solutions exactly. On the other hand, establishing the accuracy and exactness of boundary schemes for the LBM requires a multiscale analysis to link the mesoscopic populations to the macroscopic solution (see the appendix). In the Chapman-Enskog expansion, the distribution function is decomposed as

$$f_i = f_i^{eq} + f_i^{(1)} + f_i^{(2)} + \cdots \tag{43}$$

This is an expansion in powers of the Knudsen number *Kn*, with $f_i^{(k)} / f_i^{eq} = O(Kn^k)$, but also in derivatives, with $f_i^{(k)} \sim (\vec{c}_i \cdot \nabla)^k f_i^{eq}$ [23]. Achieving second-order accuracy at the boundary requires computing the first two terms in (43) accurately, which is what the regularized schemes do. However, keeping only $f_i^{(1)}$ neglects second order derivatives, so the scheme cannot reproduce parabolic solutions exactly, which would require modelling $f_i^{(2)}$ as well. On the other hand, the bounce-back scheme does no truncate the populations but merely reflects them (and adds in a small correction to properly reconstruct the velocity gradient). The simplified Chapman-Enskog analysis discussed in section 3.2.2 and in the appendix shows that the bounce-back rule may be expected to reproduce the behavior of the velocity up to second-order, which may also explain the different behaviors observed.

## 5. Conclusions

In this work, the formulation of wall boundary conditions for LBM coupled to the Spalart-Allmaras turbulence model using wall functions has been investigated. While the scheme does not solve the RANS equations, LBM computations converge to RANS-like solutions statistically. Our analysis has focused on two widely used boundary schemes: the regularized and the bounce-back formulations. The baseline regularized scheme is the one recently introduced by Cai et al. [20, 21] for high Reynolds number turbulent flows in the LBM-RANS context. A hybrid regularized scheme is introduced as a tool to analyze the influence of the wall-normal velocity gradient reconstruction involved in estimating the non-equilibrium distribution functions within the regularized approach. This method blends second-order accurate finite difference approximations from the flow field with analytical derivatives obtained from the wall function. Additionally, an alternative boundary scheme based on the slip-velocity bounce-back formulation has been adapted for LBM simulations with wall functions.

The performance of the three wall boundary schemes has been analyzed in two canonical test cases: turbulent channel flow and zero-pressure-gradient flat plate. These cases were selected to isolate the influence of boundary condition treatments, excluding other error sources such as interpolation inaccuracies or pressure gradient effects due to surface curvature. The results show that, in these simple test cases, the slip-velocity bounce-back scheme behaves fairly consistently in terms of both accuracy and sensitivity to mesh resolution. The regularized-based approach, on the other hand, appears to be highly sensitive to mesh density, even in simple geometries such as flat walls where no interpolation is required. The reason for that is likely due to the fact that the regularized boundary condition requires the numerical computation of near-wall velocity gradients, which is difficult in under-resolved grids as has been noted before in the literature. The hybrid regularized scheme addresses this issue by blending numerical and analytic gradient information, while the method of Cai et al. introduces a specially designed third-order differentiation operator. The bounce-back method, on the other hand, does not require gradient information, but careful interpolation is necessary when the wall is not placed half-way between mesh nodes, which may complicate matters in curved geometries.

The hybrid regularized scheme requires *a priori* calibration, which clearly limits its applicability to more general flow scenarios. Calibration is performed in this paper by matching a reference canonical flow solution (the turbulent channel flow). The same calibration yields reasonable results in the flat plate case, but we do not expect this behavior to hold in more general scenarios, thus limiting its potential applicability to more complex cases, where re-calibration using other reference flow solutions should be carried out. At any rate, the hybrid scheme is introduced more as a tool to investigate the influence of gradient reconstruction schemes on the regularized boundary condition than as a feasible boundary scheme on its own.

The bounce-back approach, on the other hand, appears to produce reliable results for very coarse meshes, while also exhibiting smooth convergence towards fully resolved RANS solutions. In the flat plate test case, the error in the skin friction coefficient remains below 5% for the coarsest grid resolution, with fewer than 10 points within the boundary layer. These conclusions are of course limited to the simple cases addressed and cannot be extrapolated to more complex configurations, where wall curvature or complex flow features such as adverse pressure gradients will play a significant role. Additional studies involving these scenarios are due, and will be pursued in future work. Anyhow, if similar conclusions held in those configurations, it would represent a significant advantage for practical 3D simulations, where the LBM method is particularly demanding in terms of grid resolution due to its isotropic discretization. Moreover, the interpolated bounce-back approach is computationally simpler and more compact than the regularized-based methods. This makes it an attractive alternative for simulations involving complex 3D geometries, especially when leveraging specialized parallelization strategies for many-core architectures in lattice Boltzmann simulations [49].

One final note concerns the dependence of the results on the turbulence model and wall function scheme used in the computations. Analogous analysis, which are not reported here, have been carried out with a different turbulence model and wall function with similar results.

Future work will focus on extending this study to curved surfaces, which inherently requires the use of immersed boundary methods in conjunction with Cartesian grids. This extension will demand the development of additional techniques to address interpolation challenges and the non-uniform placement of boundary points within the turbulent boundary layer.

## Appendix: Simplified Chapman-Enskog analysis of the slip velocity bounce back scheme

In this appendix we will try to derive the macroscopic behavior of the slip velocity bounce back rule using Chapman-Enskog analysis. We mainly follow the arguments in [23] adapted to the present case. We will use several simplifications:

1. BGK collision model

$$f_i(\vec{x}+\vec{c}_i,t+1)=f_i(\vec{x},t)-\tau^{-1}(f_i(\vec{x},t)-f_i^{eq}(\vec{x},t))=f_i^{eq}(\vec{x},t)+(1-\Delta t/\overline{\tau})f_i^{neq}(\vec{x},t) \quad (44)$$

where $f_i^{neq}(\vec{x},t)=f_i(\vec{x},t)-f_i^{eq}(\vec{x},t)$.

2. 2D case, D2Q9 lattice with velocities

$$\begin{aligned} c_{ix} &= c(0,0,1,1,1,0,-1,-1,-1) \\ c_{iy} &= c(0,1,1,0,-1,-1,-1,0,1) \end{aligned} \quad (45)$$

where $c=\Delta x/\Delta t$.

3. Linearized equilibrium distributions

$$f_i^{eq}(\vec{x},t)\approx w_i(\rho+\rho_0\vec{u}\cdot\vec{c}_i/\overline{c}_s^2) \quad (46)$$

where $\rho_0$ is the stagnation density and $\overline{c}_s=cc_s$ is the dimensional speed of sound.

4. Steady state, $\partial_t f_i(\vec{x},t)=0$

The Chapman-Enskog analysis of the Lattice-Boltzmann equations starts by considering that the distribution functions can be expanded in powers of the Knudsen number *Kn*, assumed small, around the equilibrium distribution

$$f_i = f_i^{eq} + \underbrace{f_i^{(1)} + f_i^{(2)} + \cdots}_{f_i^{neq}} \tag{47}$$

where $f_i^{(k)} / f_i^{eq} = O(Kn^k)$ . Introducing this expansion in the Lattice-Bolzmann equation and equating equal powers of *Kn* yields, at the first two orders, the following representation [23]

$$\begin{aligned} f_i^{(1)} &= -\bar{\tau}\vec{c}_i \cdot \nabla f_i^{eq} \\ f_i^{(2)} &= \bar{\tau}(\bar{\tau} - \Delta t / 2)(\vec{c}_i \cdot \nabla)^2 f_i^{eq} \end{aligned} \tag{48}$$

Under these assumptions, the second-order Chapman-Enskog expansion of the bounce-back formula eq. (34) reads, after using eq. (47) into the post-streaming $f_{\hat{i}}(\vec{x}_B, t+1)$ and post-collision populations $f_i^*(\vec{x}_B, t) = f_i^{eq}(\vec{x}_B, t) + (1 - \bar{\tau}^{-1}\Delta t) f_i^{neq}(\vec{x}_B, t)$ , as follows:

$$f_{\hat{i}}^{eq} + f_{\hat{i}}^{(1)} + f_{\hat{i}}^{(2)} = f_i^{eq} + (1 - \bar{\tau}^{-1}\Delta t)(f_i^{(1)} + f_i^{(2)}) - 2 w_i \rho_0 \frac{\vec{c}_i \cdot \vec{u}_w}{\bar{c}_s^2} \tag{49}$$

Inserting eq. (48) into eq. (49), using eq. (46), the equation of state $p = \rho \bar{c}_s^2$ and the identities $w_{\hat{i}} = w_i$ and $\vec{c}_{\hat{i}} = -\vec{c}_i$ , and rearranging, yields

$$\begin{aligned} \vec{c}_i \cdot \vec{u}_w = \vec{u} \cdot \vec{c}_i + \Delta t \vec{c}_i \cdot \nabla(\vec{u} \cdot \vec{c}_i) / 2 + (\bar{\tau} - \Delta t / 2)^2 (\vec{c}_i \cdot \nabla)^2 (\vec{u} \cdot \vec{c}_i) \\ - \rho_0^{-1} (\bar{\tau} - \Delta t / 2)\left(\vec{c}_i \cdot \nabla p - \Delta t (\vec{c}_i \cdot \nabla)^2 p / 2\right) \end{aligned} \tag{50}$$

We now assume that the wall lies along the *x* axis and that the velocity only varies significantly along the normal (positive *y*) direction, while $\partial_y p \approx 0$ . Then only diagonal links (those with $c_{ix} = \pm c_{iy} \neq 0$ ) play a role and eq. (50) yields

$$u_w = u_B + \Delta t c_{iy} \partial_y u / 2 + (\bar{\tau} - \Delta t / 2)^2 (c_{iy})^2 \partial_y^2 u - \rho_0^{-1} (\bar{\tau} - \Delta t / 2)\left(\partial_x p - \Delta t c_{ix} \partial_x^2 p / 2\right) \tag{51}$$

Using $\Delta t c_{iy} = -\Delta x$ and considering pressure-driven flows obeying $\rho_0^{-1} \partial_x p = \nu \partial_y^2 u$ , with $\nu = \bar{c}_s^2 \left(\bar{\tau} - \Delta t / 2\right)$ , and $\partial_x^2 p = 0$ , eq. (51) reads

$$u_w = u_B + \Delta t c_{iy} \partial_y u / 2 + c^2 (\bar{\tau} - \Delta t / 2)^2 (1 - c_s^2) \partial_y^2 u = u_B - \frac{1}{2} \Delta x \partial_y u + \frac{2}{3} \Delta x^2 (\tau - 1/2)^2 \partial_y^2 u \tag{52}$$

which is eq. (37).

## Funding

The research described in this paper has been supported by INTA and the Ministry of Defence of Spain under the grant IDATEC (IGB21001).

## Data Availability

The data that support the findings of this study are available from the corresponding authors upon reasonable request.